%% file: triple_soft.tex
\documentclass[12pt]{article}
\usepackage{jheppub}
\pdfoutput=1
\usepackage{amsmath,amssymb}
\usepackage{mathtools}
\usepackage{amsthm}
\usepackage{graphicx}
\usepackage[export]{adjustbox}
\usepackage[utf8]{inputenc}
\usepackage{slashed}
\usepackage{bbold}
\usepackage{tikz}
\usepackage{braket}
\usetikzlibrary{arrows,shapes}
\usepackage{float}
\usepackage{subfig}
\usepackage[most]{tcolorbox}
\usepackage{pdfpages}
\usepackage[mathscr]{euscript}
\usepackage{hyperref}
\usepackage{float}
\usepackage{amsfonts, bm}
\usepackage{mathrsfs}
\usepackage[numbers]{natbib}
\usepackage{url}

\def\beq{\begin{equation}}
\def\eeq{\end{equation}}
\def\beqa{\begin{eqnarray}}
\def\eeqa{\end{eqnarray}}
\newcommand{\nn}{\nonumber}

\def\eqn#1{eq.~(\ref{#1})}

\DeclareMathOperator{\Tr}{Tr}
\def\bea{\begin{eqnarray}}
\def\eea{\end{eqnarray}}
\newcommand{\rayan}[1]{{\color{red}RH: #1}}

\newcommand{\Eik}[2]{\mathcal{S}_{#1}(p_{#2})}
\newcommand{\K}[1]{k_{\perp #1}}
\newcommand{\pM}[1]{{P}^{#1}}
\newcommand{\KT}[1]{\tilde{k}_{\perp #1}}
\def\eps{\epsilon}
\newcommand{\Kb}[1]{\kappa_{\perp #1}}
\newcommand{\KTb}[1]{\tilde{\kappa}_{\perp #1}}
\newcommand{\pN}[1]{\widetilde{P}'^{#1}}
\def\cM{{\cal M}}  
  
\def\cS{{\cal S}}

\definecolor{desycyan}{rgb}{0.00,0.68,0.93}
\definecolor{desyorange}{rgb}{0.93,0.58,0.16}

\newcommand\numberthis{\addtocounter{equation}{1}\tag{\theequation}}

\numberwithin{equation}{section}

\title{Tree-level soft emission of a quark pair in association with a gluon}
\author[a,b]{Vittorio Del Duca\footnote{On leave from INFN, Laboratori Nazionali di Frascati, Italy.}}
\author[c]{Claude Duhr}
\author[a]{Rayan Haindl}
\author[d]{and Zhengwen Liu}

\affiliation[a]{Institute for Theoretical Physics, ETH Z\"{u}rich, 8093 Z\"{u}rich, Switzerland}
\affiliation[b]{Physik-Institut, Universit\"{a}t Z\"{u}rich, Winterthurerstrasse 190, 8057 Z\"{u}rich, Switzerland}
\affiliation[c]{Bethe Center for Theoretical Physics, Universit\"{a}t Bonn, 53115 Bonn, Germany}
\affiliation[d]{Deutsches Elektronen-Synchrotron DESY, Notkestr.\,85, 22607 Hamburg, Germany}

\emailAdd{delducav@itp.phys.ethz.ch}
\emailAdd{cduhr@uni-bonn.de}
\emailAdd{haindlr@phys.ethz.ch}
\emailAdd{zhengwen.liu@desy.de}

\abstract{
We compute the tree-level current for the emission of a soft quark-antiquark pair in association with a gluon. This soft current is the last missing ingredient to understand the infrared singularities that can arise in next-to-next-to-next-to-leading-order (N$^3$LO) computations in QCD. Its square allows us to understand for the first time the colour correlations induced by the soft emission of a quark pair and a gluon. We find that there are three types of correlations: Besides dipole-type correlations that have already appeared in soft limits of tree-level amplitudes, we uncover for the first time also a three-parton correlation involving a totally symmetric structure constant. We also study the behaviour of collinear splitting amplitudes in the triple-soft limit, and we derive the corresponding factorisation formula.
}

\begin{document}
\preprint{BONN-TH-2022-12, DESY-22-073}

\setcounter{tocdepth}{2}
\maketitle

\section{Introduction}

The cornerstone of all methods to make predictions for modern collider experiments is perturbative Quantum Field Theory, where observables are expanded into a series in the coupling constants. The higher orders in the perturbative expansion capture, on the one hand, the effect of the exchange of virtual quanta, and, on the other hand, the emission of unobserved real particles in the final state. In theories featuring massless particles, both the real and virtual corrections are separately divergent (even after ultraviolet renormalisation), but their sum is finite for a scattering of colourless particles in the initial state (e.g., at an $e^+e^-$ collider) due to the celebrated Kinoshita-Lee-Nauenberg theorem.\footnote{If the scattering features also hadrons in the initial state, the divergences only cancel after we perform mass factorisation.} The cancellation of these infrared singularities, however, is very intricate, because it only happens after phase space integration, and the real and virtual corrections live in different phase spaces. 

The infrared singularities stemming from real corrections arise from regions of phase space where massless particles become unresolved, i.e., they become either soft (meaning that they have vanishing energies) or collinear to each other. The behaviour of scattering amplitudes in these unresolved limits is universal, in the sense that the amplitudes factorise into amplitudes without the unresolved particles, multiplied by a function that captures the divergence and does not depend on the underlying hard scattering. This universality of soft and collinear divergences is at the heart of so-called subtraction schemes, where, very loosely speaking, one subtracts the phase space divergences of real corrections at the integrand level, and adds them back in integrated form to cancel the infrared singularities of virtual corrections. Subtraction schemes at next-to-leading order (NLO) and next-to-next-to-leading order (NNLO) in the strong coupling constant are one of the cornerstones of modern precision computations in (massless) Quantum Chromodynamics (QCD). The construction and the success of subtraction schemes at NLO and NNLO rely crucially on the fact that the soft and collinear divergences describing tree-level amplitudes with up to two unresolved particles and one-loop amplitudes with one unresolved parton are well understood~\cite{Campbell:1997hg,Catani:1998nv,Catani:1999ss,DelDuca:1999iql,Kosower:2002su,Czakon:2011ve,Bern:1994zx,Bern:1998sc,Kosower:1999rx,Bern:1999ry,Catani:2000pi,Braun-White:2022rtg}.

In order to reach the target precision for current and future collider experiments, like the Large Hadron Collider (LHC) at CERN and its potential successors, NNLO computations in QCD may not be sufficient, but also next-to-next-to-next-to-leading order (N$^3$LO) corrections will be required. While first examples of non-trivial two- and three-loop amplitudes relevant to N$^3$LO computations have recently become available~\cite{Caola:2021rqz,Bargiela:2021wuy,Caola:2021izf,Abreu:2021oya,Abreu:2021asb,Badger:2021nhg,Badger:2021ega,Badger:2022ncb}, one of the major bottlenecks is that there is currently no general understanding of how to combine the real and virtual corrections. Experience from NNLO shows that it is important to understand in detail all the unresolved limits that lead to singularities in an N$^3$LO computation. The relevant collinear singularities are by now completely understood, and they include the emission of four collinear partons at tree-level ~\cite{DelDuca:2019ggv,DelDuca:2020vst},
three collinear partons at one loop~\cite{Catani:2003vu,Badger:2015cxa,Czakon:2022fqi} or two collinear partons at two loops~\cite{Duhr:2014nda}. Soft singularities at N$^3$LO have also been studied. In particular, the emission of a soft gluon at two loops is well established~\cite{Li:2013lsa,Duhr:2013msa,Dixon:2019lnw}, as is the emission of a pair of soft gluons or quarks at one loop~\cite{Zhu:2020ftr,Catani:2021kcy}. Tree-level soft emission of three partons, however, has so far only been studied for the emission of three soft gluons~\cite{Catani:2019nqv}, but the case of a soft quark pair in addition to a gluon at tree-level has never been considered. The main goal of this paper is to provide for the first time this last soft limit needed to describe all infrared singularities that can arise in N$^3$LO computations.

%from the emission of a single-soft gluon at two-loops
%or three soft gluons~\cite{Catani:2019nqv}, one-loop amplitudes with either three collinear partons~\cite{Catani:2003vu,Badger:2015cxa,Czakon:2022fqi} or two soft partons~\cite{Zhu:2020ftr,Catani:2021kcy}, and two-loop amplitudes with either two collinear partons~\cite{Bern:2004cz,Badger:2004uk,Duhr:2014nda} or one soft gluon~\cite{Li:2013lsa,Duhr:2013msa,Dixon:2019lnw} have been computed. 

%The only missing piece of the list above of IR currents at ${\rm N^3LO}$ accuracy is the tree-level emission of 
%a soft quark-antiquark pair and a soft gluon, which yields a leading divergence in the phase-space region with three unresolved partons.
%In this work, we present it for the first time.

The paper is organised as follows: In section~\ref{sec:soft_currents},
we review soft limits of tree-level scattering amplitudes, displaying explicitly the currents for a soft gluon and a soft $\bar{q}q$ pair.
In section~\ref{sec:triple_soft}, we present the main results of this paper, namely the tree-level current for the soft $\bar{q}qg$ emission from QCD scattering amplitudes. In section~\ref{sec:kinematic}, we consider kinematic sub-limits of the soft $\bar{q}qg$ limit, which 
are useful for constructing subtraction schemes at N$^3$LO accuracy. In section~\ref{sec:conclusion} we draw our conclusions.
We also include two appendices where we display analytic results which are too lengthy to be shown in the main text.

\section{Tree-level soft currents}
\label{sec:soft_currents}

The aim of this paper is to study the behavior of tree-level QCD amplitudes in the limit where a certain number of massless partons are \emph{soft}, i.e., they have vanishing energies. To be more precise, consider the scattering of $n$ particles with momenta $p_i$, and flavour, helicity and colour indices $f_i$, $s_i$ and $c_i$, respectively.
%Tree-level QCD amplitudes factorise in the limit where a number of partons have vanishing momenta.
If a subset of $m$ massless partons become soft, the amplitude is divergent and the leading behaviour is captured by the factorisation formula,
\begin{align}
\begin{split}
    &\mathscr{S}_{1\ldots m}\,\mathcal{M}^{c_1\ldots c_n;s_{1}\ldots s_{n}}_{f_1\ldots f_n}(p_{1},\ldots,p_n)
    \\&\qquad \qquad= (\mu^{\eps}g_s)^m\, \mathbf{J}^{c_1\ldots c_m;s_1\ldots s_m}_{f_1\ldots f_m}(p_1,\ldots,p_m)\, \mathcal{M}^{c_{m+1}\ldots c_n;s_{m+1}\ldots s_{n}}_{f_{m+1}\ldots f_n}(p_{m+1},\ldots,p_n)\,,
\end{split}
    \label{eq:factorisation_multisoft}
\end{align}
where $g_s$ is the strong coupling constant and $\mu$ is the scale introduced by dimensional regularisation. Throughout this paper we work in Conventional Dimensional Regularisation (CDR) in $D=4-2\epsilon$ dimensions. In particular, we assume that gluons have $D-2$ physical polarisations (quarks always have 2 polarisations). The symbol $\mathscr{S}_{1\ldots m}$ in eq.~\eqref{eq:factorisation_multisoft} denotes the operation of keeping only the leading divergent term in the soft limit.
The scattering amplitude $\mathcal{M}^{c_{m+1}\ldots c_n;s_{m+1}\ldots s_{n}}_{f_{m+1}\ldots f_n}$ on the right-hand side is obtained from the amplitude on the left-hand side by simply removing the soft particles. The current $\mathbf{J}^{c_1\ldots c_m;s_1\ldots s_m}_{f_1\ldots f_m}$ describes the leading divergent behaviour of the amplitude in the soft limit, often referred to as the \emph{eikonal} approximation in the literature. 

The soft current depends on the colour, spin and flavour quantum numbers of the soft partons.
In order to keep our notations compact, we find it useful to work with the colour-space formalism of refs.~\cite{Catani:1996jh,Catani:1996vz}, where scattering amplitudes are interpreted as vectors that can be expanded into an orthonormal basis in colour $\otimes$ spin space,
\begin{equation}
    |\cM_{f_1,\ldots f_n}(p_1,\ldots,p_n)\rangle = |c_1\ldots c_n\rangle\otimes |s_1\ldots s_n\rangle\,\mathcal{M}^{c_1\ldots c_n;s_{1}\ldots s_{n}}_{f_1\ldots f_n}(p_1,\ldots,p_n)\,.
\end{equation}
We will often suppress the dependence on the momenta. With this notation, the squared matrix element summed over spin and colour indices of the external particles can be written as
\begin{align}
\begin{split}
\left|\cM_{f_1\ldots f_n}\right|^2 
\equiv \langle\cM_{f_1\ldots f_n} | \cM_{f_1\ldots f_n} \rangle  
&=
\sum_{\substack{(s_1,\ldots,s_n)\\(c_1,\ldots,c_n)}}\left[\cM_{f_1\ldots f_n}^{c_1\ldots c_n;s_1\ldots s_n}\right]^\dagger
\cM_{f_1\ldots f_n}^{c_1\ldots c_n;s_1\ldots s_n}
\,.
\end{split}
\label{eq:ampl_squared}
\end{align}
The soft current can then be interpreted as an operator $\mathbf{J}_{f_1\ldots f_m}(p_1,\dots,p_m)$ in this colour space, and the soft factorisation in eq.\,\eqref{eq:factorisation_multisoft} takes the form,
\begin{align}
    \begin{split}
    &\mathscr{S}_{1\ldots m}\,
    |\cM_{f_1\ldots f_n}\rangle 
    =
    (\mu^{\eps}g_s)^m\, \mathbf{J}_{f_1\ldots f_m}\,
    |\cM_{f_{m+1}\ldots f_n}\rangle\,.
    \end{split}
    \label{eq:factorisation_multisoft_CSspace}
\end{align}
This operator acts on color space via the infinitesimal generators of the gauge transformations,
\begin{equation}
    {\bf T}_i^a |c_1 \ldots c_n\rangle = |c_1 \ldots c'_i\ldots c_n\rangle\,{\bf T}_{c'_ic_i}^a\,,
\end{equation}
where we defined ${\bf T}^a_{c'_i c_i} = t^a_{c'_ic_i}$ if parton $i$ is a quark, ${\bf T}^a_{c'_i c_i} = - t^a_{c_ic'_i}$ if it is an anti-quark (here $t^a_{c'_ic_i}$ are the generators of the fundamental representation of SU$(N_c)$), and ${\bf T}^a_{c'_i c_i} = i f^{c'_iac_i}$ for a gluon. 
Since colour must be conserved in every scattering process, the vector $|\cM_{f_{m+1},\ldots,f_n}\rangle$ must be a colour singlet, i.e., it must satisfy
\begin{align}
    \sum_{i=m+1}^n \mathbf{T}^a_i |\cM_{f_{m+1}\ldots f_m}\rangle =0\,.
\end{align}
Henceforth we will simply use the shorthand,
\begin{align}
    \sum_{i=m+1}^n \mathbf{T}_i^a =0\,,
\end{align}
where it is understood that this identity is only valid when we act on colour singlet states. 

The soft current for the emission of a single soft gluon is known through two loops in the strong coupling constant~\cite{BASSETTO1983201,Proceedings:1992fla,Bern:1998sc,Bern:1999ry,Catani:2000pi,Duhr:2013msa,Li:2013lsa,Dixon:2019lnw}. The double-soft current is known at tree-level and one-loop for the emission of a pair of soft gluons or quarks~\cite{Campbell:1997hg,Catani:1999ss,Czakon:2011ve,Zhu:2020ftr,Catani:2021kcy}. Triple-soft emission is currently only known at tree-level for three gluons~\cite{Catani:2019nqv}.\footnote{Quadruple-soft emission is known at tree-level for four gluons in the special case of the emission from two hard partons~\cite{Catani:2019nqv}.} The main aim of this paper is to compute for the first time the triple soft current for the emission of a soft quark pair in addition to a gluon. In the remainder of this section, we review the known results for the tree-level soft currents for the emission of a single soft gluon or quark pair, both in order to illustrate some general properties of soft currents and to define some quantities that will be useful when computing the soft current ${\bf J}_{\bar{q}qg}$ in section~\ref{sec:triple_soft}. 

\subsection{The tree-level soft current for the emission of a single soft gluon}
\label{sec:J_g_review}
Let us start by discussing the case of the emission of a single soft gluon at tree-level. It will often be useful to consider a variant of the soft current operator where we have amputated the polarisation states, e.g., in the case of a single soft gluon, we define
\begin{equation}
    {\bf J}_g(p_1) = |a\rangle\otimes|s\rangle\,{\bf J}^{a;s}_g(p_1)=|a\rangle\otimes|s\rangle\,\epsilon^s_\mu(p_1,n)\,{\bf J}^{a;\mu}_g(p_1)\,.
\end{equation}
Note that throughout this paper we always work in axial gauge for external gluons, where the polarisation vectors satisfy the constraints,
\begin{equation}
p_1^\mu \epsilon_\mu^s(p_1,n)=n^\mu \epsilon_\mu^s(p_1,n)=0 \,,
\label{eq:axgauge}
\end{equation}
where $n$ is a lightlike reference vector with $p_1\cdot n\neq0$.
The soft current for the tree-level emission of a single gluon is given by~\cite{BASSETTO1983201,Proceedings:1992fla}
\begin{align}
    \mathbf{J}^{a;\mu}_g(p_1) = 
    \sum_{i=2}^n \mathbf{J}^{a;\mu}_i(p_1) =
    \sum_{i=2}^n
    \mathcal{S}_i^\mu(p_1)\mathbf{T}_i^{a}\,,
    \label{eq:single_soft}
\end{align}
with
\begin{align}
    \mathcal{S}_i^\mu(p_1) = \frac{p_i^\mu}{p_i\cdot p_1}\,.
    \label{eq:eikonal}
\end{align}
Note that, as a consequence of gauge invariance, the soft current is conserved, 
\begin{equation}
\label{eq:div_J_g}
    p_{1\mu}\,\mathbf{J}^{a;\mu}_g(p_1) = \sum_{i=2}^n
    \mathbf{T}_i^{a} = 0 \mod \sum_{i=2}^n\mathbf{T}_i^{a} =0\,.
\end{equation}
A similar relation also holds for multi-soft gluon emission, albeit, due to the non-abelian nature of the gauge interactions, only upon replacing \emph{one} polarisation vector by its momentum. In ref.~\cite{Catani:2019nqv} a stronger version was shown to hold even if the remaining external polarisation vectors are amputated, although one needs to project onto the subspace spanned by the physical polarisations. Since our main interest is the current for the emission of a single soft gluon  in addition to a soft quark pair, we will not pursue this further.

In applications one is usually interested in the effect of the soft current and the color-correlations it induces after squaring the matrix element,
    \begin{align}
\begin{split}
    \mathscr{S}_{1}\,
   %\langle 1,\ldots,n|1,\ldots,n\rangle 
   |\cM_{gf_2\ldots f_n}|^2
 &\,=
   (\mu^{\eps}g_s)^{2}
   \langle \cM_{f_{2}\ldots f_n}|
   |\mathbf{J}_{g}(p_1)|^2|\cM_{f_{2}\ldots f_n}\rangle\\
  &\,    = -(\mu^{\eps}g_s)^{2}
    \sum_{i,j=2}^n \mathcal{S}_{ij}(p_1)
    |\mathcal{M}_{f_2\ldots f_n}^{(ij)}|^2
    \label{eq:single_soft_fact}\,,
\end{split}
\end{align}
where we have introduced the colour-correlated matrix element,
\begin{equation}\label{eq:dipole_M_ij}
    |\mathcal{M}_{f_2\ldots f_n}^{(ij)}|^2 = \langle \cM_{f_{2}\ldots f_n}|
   \mathbf{T}_i\cdot \mathbf{T}_j|\cM_{f_{2}\ldots f_n}\rangle\,,
\end{equation}
and the well-known eikonal function,
\begin{align}
    \mathcal{S}_{ij}(p_1) = \frac{2\chi_{ij}}{\chi_{1i}\chi_{1j}}
    %=\frac{p_i\cdotnew p_j}{(p_1\cdotnew p_i)( p_1\cdotnew p_j)}\,,
    \label{eq:eik_g}
\end{align}
with $\chi_{ij}=2p_i\cdot p_j$ and $\mathbf{T}_i\cdot \mathbf{T}_j =  \mathbf{T}_j\cdot \mathbf{T}_i =  \mathbf{T}_i^a \mathbf{T}_j^a$. At this point we make some important comments. First, if we simply evaluate the square of the soft current in axial gauge, we find
\begin{align}
\label{eq:singlsquaresoft}
    |\mathbf{J}_g(p_1)|^2 
    & = \left[\mathbf{J}^{a;\mu}_g(p_1)\right]^\dagger
    d_{\mu\nu}(p_1,n)\mathbf{J}^{a;\nu}_g(p_1) \\
    = & - \sum_{i,j=2}^n \,\mathbf{T}_i \cdot \mathbf{T}_j \, \mathcal{S}_{ij}(p_1) +\sum_{i,j=2}^n\,\mathbf{T}_i \cdot \mathbf{T}_j \, 
    \left(\frac{n\cdot p_i}{(p_i\cdot p_1) (n \cdot p_1)}
    +\frac{n\cdot p_j}{(p_j\cdot p_1) (n \cdot p_1)}
    \right)\,,
\nonumber
\end{align}
where $d_{\mu\nu}(p_1,n)$ denotes the sum over physical polarisations in axial gauge,
\begin{align}
   d_{\mu\nu}(p_1,n) = \sum_{s=1}^{D-2} \epsilon^{s}_{\mu}(p_1,n)\epsilon^{s}_{\nu}(p_1,n)= -g_{\mu\nu}+\frac{p_{1\mu} n_\nu+n_\mu p_{1\nu}}{p_1\cdot n}\,.
    \label{eq:polarisation_sum}
\end{align}
We see that the terms dependent on $n$ cancel due to colour conservation when acting on the amplitude in eq.~\eqref{eq:single_soft_fact}. We will from now on drop terms in the squared current that vanish due to colour conservation in the hard amplitude. Second, we note that, as a consequence of $\mathbf{T}_i\cdot \mathbf{T}_j =  \mathbf{T}_j\cdot \mathbf{T}_i$, the colour-correlated squared amplitude, $|\mathcal{M}_{f_2\ldots f_n}^{(ij)}|^2$,
%eikonal function, eq.~\eqref{eq:eik_g}, 
features a symmetry under exchange of the emitters $(i\leftrightarrow j)$, to which we will refer as the \textit{dipole symmetry}. The dipole symmetry of $|\mathcal{M}_{f_2\ldots f_n}^{(ij)}|^2$ implies that the kinematic factor (the eikonal function $\cS_{ij}(p_1)$) in \eqn{eq:single_soft_fact} is symmetric as well under the exchange of the emitters, $(i\leftrightarrow j)$. This is a general feature: The kinematic functions in the squared soft current inherit the symmetry properties of the colour operators that they multiply.

\subsection{The tree-level soft current for the emission of a single quark pair}
Let us also briefly review the features of the soft current for a single soft quark pair emission. It is again convenient to amputate the polarisation states,
\begin{equation}\begin{split}
    \mathbf{J}_{\bar{q}q}(p_1,p_2) &\,= |\bar{\imath}j\rangle\otimes|s_1s_2\rangle\,
    \mathbf{J}_{\bar{q}q}^{\bar{\imath}j;s_1s_2}(p_1,p_2)\\
    &\,= |\bar{\imath}j\rangle\otimes|s_1s_2\rangle\,
    \overline{u}_{s_2}(p_2)\mathbf{J}_{\bar{q}q}^{\bar{\imath}j}(p_1,p_2)v_{s_1}(p_1)\,,
\end{split}\end{equation}
where $\bar{\imath}$ and $j$ denote the anti-fundamental and fundamental colour indices of the anti-quark and quark respectively.
Note that $\mathbf{J}_{\bar{q}q}^{\bar{\imath}j}$ is a matrix in Dirac spinor space, though we suppress the matrix indices for brevity. It is easy to see that the current for the emission of a soft quark pair is entirely determined by the soft current for single-gluon emission,
\begin{equation}
\mathbf{J}_{\bar{q}q}^{\bar{\imath}j}(p_1,p_2) = -\frac{1}{\chi_{12}}\,t^a_{j\bar{\imath}}\,\gamma_{\mu}\,\mathbf{J}^{a;\mu}_{g}(p_{12})\,,\qquad p_{12} = p_1+p_2\,,
\label{eq:from_qqbar_to_g}
\eeq
where we work with a gluon propagator in Feynman gauge.
We can square the soft current to obtain
\begin{align}
\begin{split}
    &\mathscr{S}_{12}
    |\cM_{\bar{q}_1{q}_2f_3\ldots f_n}|^2
    %|&\mathcal{M}_{g_1f_2\ldots f_n}|^2
    %\\&
    = (\mu^{2\eps}g_s^2)^2\,T_F\,\sum_{i,j=3}^nQ^{\bar{q}q}_{ij}(p_1,p_2)
    |\mathcal{M}_{f_3\ldots f_n}^{(ij)}|^2 \,,
    \label{eq:qq_soft_fact}
\end{split}
\end{align}
where $T_F=\frac{1}{2}$ is defined by $\textrm{Tr}(t^at^b) = T_F\delta^{ab}$, and we defined the kinematic function,
\begin{align}
    Q^{\bar{q}q}_{ij}(p_1,p_2) = \frac{4}{\chi_{12}^2}\frac{\chi_{1i}\chi_{2j}+\chi_{1j}\chi_{2i}-\chi_{ij}\chi_{12}}{\chi_{i(12)}\chi_{j(12)}}\,,
    \label{eq:eik_qq}
\end{align}
and $\chi_{i(jk)}= \chi_{ij}+\chi_{ik}$.
Note that, in addition to the dipole symmetry under exchange of the emitters $(i\leftrightarrow j)$, the soft $\bar qq$ function features charge conjugation symmetry under the exchange of the quark and antiquark labels,
\beq
Q^{\bar{q}q}_{ij}(p_1,p_2) = Q^{\bar{q}q}_{ij}(p_2,p_1) \,.
\eeq

Finally, we recall that the eikonal currents, and so the kinematic functions of a single soft-gluon \eqref{eq:eik_g} or of a soft $\bar{q}q$ pair \eqref{eq:eik_qq}, do not depend on the mass of the emitters, while the eikonal current of two soft gluons receives an extra contribution which is proportional to the squared mass of the  emitters~\cite{Czakon:2011ve}.

%%%%%%%%%%%%%%%%%%%%%%%%%%%%%%%%%%%%%%%%%%%%%%%%%%%%
%%%%%%%%%%%%%%%%%%%%%%%%%%%%%%%%%%%%%%%%%%%%%%%%%%%%

\section{Tree-level factorisation for soft $\bar{q}qg$ emission}
\label{sec:triple_soft}

In this section we present the main result of this paper, namely our result for the tree-level soft current for soft  $\bar{q}qg$ emission. We start by quoting the results for the soft current, and then we discuss the analytic expressions for the squared current and the ensuing colour correlations.

\begin{figure}[h!]
    \centering
    \renewcommand\thesubfigure{\Roman{subfigure}}
    \subfloat[%Soft gluon emission from a soft $q\bar{q}$ pair
    ]{\includegraphics[width= 5cm]{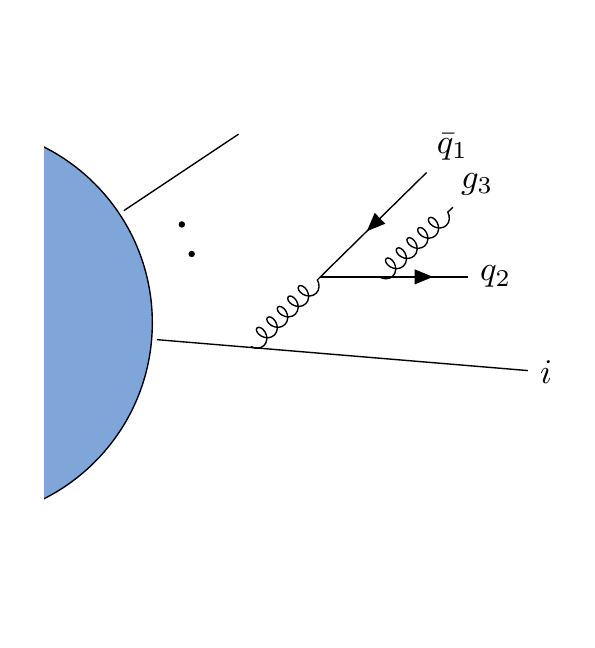}\includegraphics[width= 5cm]{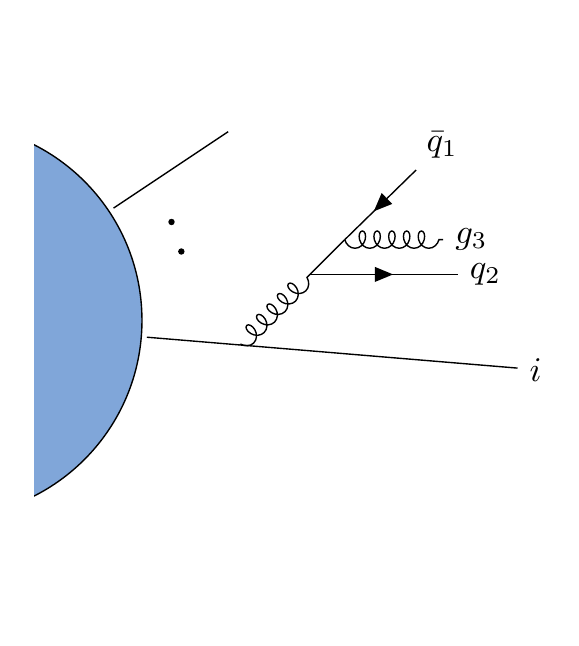}\label{subfig:qqgdiag1}}
    \hfill
    \subfloat[]{\includegraphics[width= 5cm]{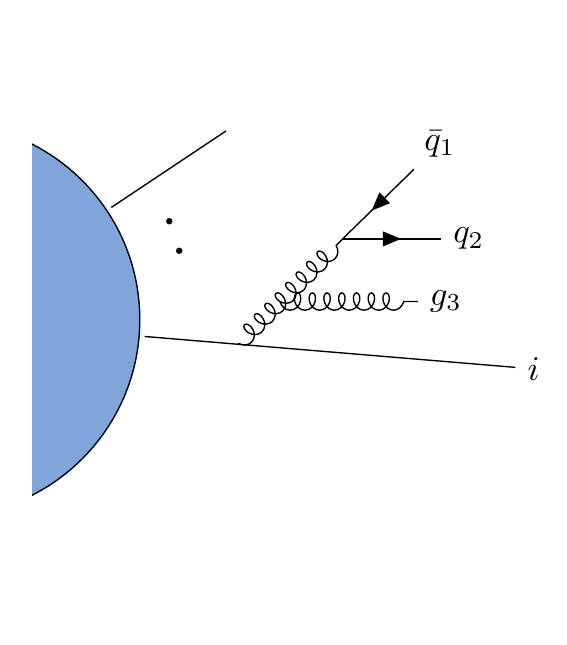}\label{subfig:qqgdiag2}}
    \hfill
    \subfloat[]{\includegraphics[width= 5cm]{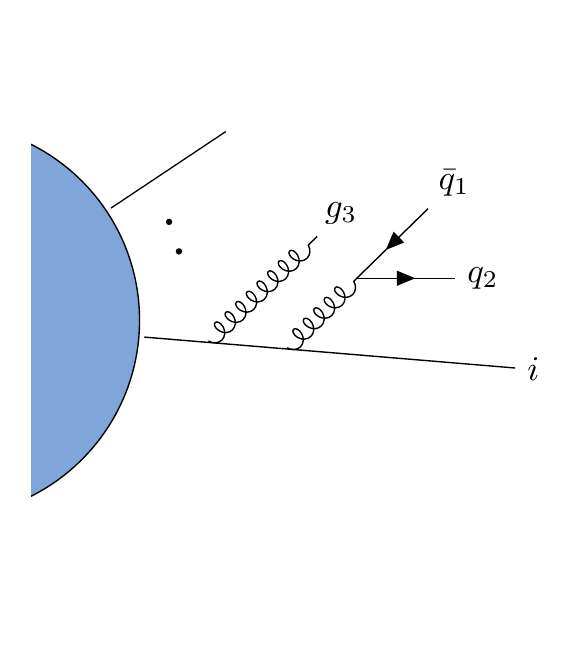}\includegraphics[width= 5cm]{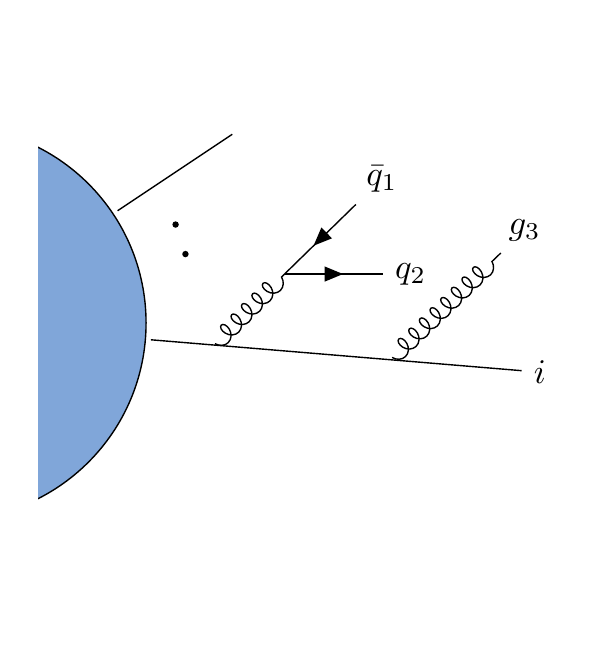}\label{subfig:qqgdiag3}}
    \subfloat[]{\includegraphics[width= 5cm]{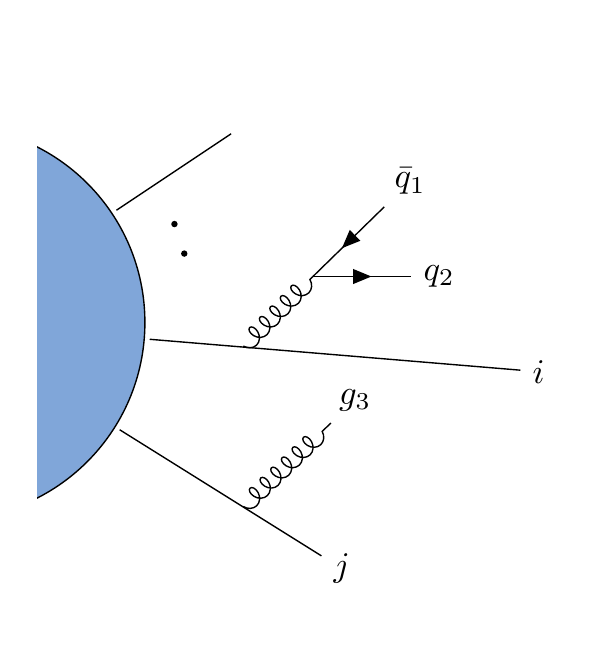}\label{subfig:qqgdiag4}}
    \caption{Soft-gluon insertion diagrams for the emission of a soft gluon and $q\bar{q}$ pair.}
    \label{fig:qqg_diag}
\end{figure}

\subsection{The soft current}
Without loss of generality, we assume the anti-quark, quark and gluon to be partons $1$, $2$ and $3$, respectively. The soft current is obtained by evaluating the diagrams in fig.\,\ref{fig:qqg_diag}, using the usual eikonal Feynman rules for the coupling of a soft gluon to a hard parton. 

We use axial gauge for the external gluon, but we evaluate all internal gluon propagators in Feynman gauge. 
We rescale the momenta of the three soft partons by $\lambda$, $p_i\to \lambda\,p_i$, $i=1,2,3$, and we perform a Laurent expansion around $\lambda=0$. The leading term in this Laurent expansion defines the soft current. We find it again convenient to amputate the polarisation vectors,
\begin{equation}
    \mathbf{J}_{\bar{q}qg}^{\bar{\imath}ja;s_1s_2s_3}(p_1,p_2,p_3) = \epsilon_{\mu}^{s_3}(p_3,n)\,\overline{u}_{s_2}(p_2)\,\mathbf{J}_{\bar{q}qg}^{a;\mu}(p_1,p_2,p_3)\,v_{s_1}(p_1)\,,
\end{equation}
where for brevity we keep the colour indices of the quark pair implicit on the right-hand side.

We split the tree-level $\bar{q}qg$ soft current into two contributions,
\begin{align}
    \mathbf{J}_{\bar{q} q g}^{a;\mu}(p_1,p_2,p_3) = 
    \mathbf{K}^{a,\mu}(p_1,p_2,p_3)
    +\mathbf{W}^{a,\mu}(p_1,p_2,p_3)\,.
    \label{eq:qqg_soft_current}
\end{align}
The first term on the right-hand side is given by,
\begin{align}
  \nonumber  %\mathbf{J}_{q\bar{q}g}^{a,\mu\,(\text{2-line})} =
    \mathbf{K}^{a,\mu}(p_1,p_2,p_3)&= 
    \frac{1}{2}\big\{\mathbf{J}_g^{a;\mu}(p_3),\mathbf{J}_{\bar{q}q}(p_1,p_2)\big\}\,,\\
 &   =- \frac{1}{2\chi_{12}} 
    \sum_{i,j =4}^n 
    %\frac{1}{2} 
    \big\{\mathbf{J}_i^{a,\mu}(p_{3}),\mathbf{J}_j^{b,\nu}(p_{12}) \big\}\gamma_\nu t^b\,.
    \label{eq:K_current} %previously 2-line current
\end{align}
where the soft currents appearing on the right-hand side have been defined in the previous section.
Equation~\eqref{eq:K_current} includes both uncorrelated ($i\neq j$) soft emission from two different legs, fig.\,\ref{subfig:qqgdiag4}, as well as contributions to the abelian part of fig.\,\ref{subfig:qqgdiag3}.
The second term on the right-hand side of eq.\,\eqref{eq:qqg_soft_current} has the explicit form,
\begin{align}
\label{eq:W_current}
     %\mathbf{J}_{q\bar{q}g}^{a,\mu\,(\text{1-line})} 
     \mathbf{W}^{a,\mu}(p_1,p_2,p_3)
     = \sum_{i=4}^n&
     \mathbf{J}_i^{b,\nu}(p_{123})\bigg\{
     \frac{1}{\chi_{123}} \left[
     \frac{1}{\chi_{13}}\gamma_\nu \slashed{p}_{13}\gamma^\mu \big(t^b t^a\big)
     - \frac{1}{\chi_{23}}\gamma^\mu \slashed{p}_{23}\gamma_\nu \big(t^a t^b\big)
     \right]
     \\
     +&
     \frac{i}{\chi_{12}}f^{abc}\left[\frac{1}{2}\delta_{\nu}^\alpha
     \left(\mathcal{S}_{i}^\mu(p_{3})-\mathcal{S}_{i}^\mu(p_{12})\right)
     + \frac{1}{\chi_{123}}U_\nu{}^{\alpha\mu}(p_{12},p_3)\right]\gamma_\alpha t^c
     \bigg\}\,,
     \nonumber
     %previously 1-line current
\end{align}
where we have defined
\begin{align}
  % \eqs[0.25]{U_vertex} \equiv
    U_\nu{}^{\alpha\mu}(p_{12},p_3) =
    - 2\delta_\nu^{\alpha}p_{12}^\mu
    + g^{\alpha\mu}(p_{12}-p_3)_\nu
    + 2\delta_{\nu}^{\mu}p_3^\alpha\,,
    \label{eq:qqg_cubic_red}
\end{align}
and 
\begin{align}
    \chi_{1\ldots k} = %\sum_{i=1}^{k-1}\sum_{\substack{j=2 \\ j>i}}^k 
    \sum_{i<j}^k
    \chi_{ij}\,.
\end{align}
Let us make two comments. First, we have seen that the soft current for the emission of a soft $\bar{q}q$ pair can be expressed as the soft current for the emission of a single soft gluon contracted with the quark current, cf.~eq.~\eqref{eq:from_qqbar_to_g}. From the Feynman diagrams in fig.~\ref{fig:qqg_diag}, it may appear that the $\bar{q}qg$ soft current can similarly be factorised into the double soft gluon current ${\bf J}_{gg}$ contracted with the quark current. This, however, is not so, because there are certain contributions to ${\bf J}_{gg}$ that can be dropped due to the transversality condition in eq.~\eqref{eq:axgauge}, but they have to be kept if the gluon is off shell. Second, we note that, as expected, we find that the divergence of the triple-soft current defined in eq.\,\eqref{eq:qqg_soft_current} is proportional to the total colour charge of the hard partons, implying that the current is conserved,
    \begin{align}
        p_{3,\mu} \mathbf{J}^{a,\mu}_{\bar{q} q g} =
        -\frac{1}{\chi_{12}}t^c\left[
        \mathbf{J}^{c,\mu}(p_{12})\gamma_\mu \delta^{ab} 
        -if^{abc}\frac{1}{\chi_{123}}\slashed{p}_3
        \right]\sum_{i=4}^n \mathbf{T}_i^b = 0 \mod \sum_{i=4}^n \mathbf{T}_i^b = 0\,,
    \label{eq:qqg_divergence}
    \end{align}
    which is reminiscent of eq.~\eqref{eq:div_J_g} in the case of single-gluon emission.

Next, we consider the structure of the squared soft current. 
We find that we can write the result as a sum over contributions from dipole, tripole and symmetrised colour correlations,
\begin{align}
    \mathscr{S}_{123}
    |\mathcal{M}_{\bar{q}_1{q}_2g_3f_4\ldots f_n}|^2  =
    \left(\mathscr{S}_{123}^{(\text{dip})}
    +\mathscr{S}_{123}^{(\text{trip})}
    +\mathscr{S}_{123}^{(\text{sym})}\right)
    |\mathcal{M}_{\bar{q}_1{q}_2g_3f_4\ldots f_n}|^2\,.
    \label{eq:qqg_soft}
\end{align}
We now discuss the three contributions in turn.
The dipole term reads
\begin{align}
\begin{split}
    &\mathscr{S}_{123}^{(\text{dip})}
    |\mathcal{M}_{\bar{q}_1{q}_2g_3f_4\ldots f_n}|^2 \\
    &\quad = 
    (\mu^{2\eps}g_s^2)^3\ T_F \sum_{i,j=4}^n 
    \left[ C_F Q_{ij}^{\bar{q}qg\,(\text{ab})} 
    + C_A \Big( Q_{ij}^{\bar{q}qg\,(\text{nab})} 
    + Q_{ij}^{\bar{q}qg\,(\text{mass})} \Big) \right]
    |\mathcal{M}^{(ij)}_{f_4\ldots f_n}|^2\,,
    \label{eq:factorisation_soft_qqg_dipole}
\end{split}
\end{align}
where $C_F$ and $C_A$ denote the Casimir operators in the fundamental and adjoint representations of SU$(N_c)$,
\begin{equation}
    C_F=\frac{N_c^2-1}{2N_c} \textrm{~~~and~~~} C_A = N_c\,.
\end{equation}
The kinematic functions will be defined below. We suppress from now on the dependence of the kinematic functions on the momenta for readability. The dipole-correlated matrix element $|\mathcal{M}^{(ij)}_{f_4\ldots f_n}|^2$ was already defined in eq.~\eqref{eq:dipole_M_ij}.
The {\it sym} term comes from the symmetrised product of two dipole operators and is given by
\begin{align}
    \begin{split}
    \mathscr{S}_{123}^{(\text{sym})}
    |\mathcal{M}_{\bar{q}_1{q}_2g_3f_4\ldots f_n}|^2 &=
    (\mu^{2\eps}g_s^2)^3\ T_F
    \sum_{i,j,k,\ell=4}^n Q_{ik;j\ell}^{\bar{q}qg}
    |\mathcal{M}^{(ik;j\ell)}_{f_4\ldots f_n}|^2\,,
    \end{split}
    \label{eq:factorisation_soft_qqg_quad}
\end{align}
where we introduced the symmetric four-parton correlation function, which also appears in the squared current for double-soft gluon emission at tree-level~\cite{Catani:1999ss},
\begin{equation}\label{eq:quadrupole}
     \big|\mathcal{M}^{(ik;j\ell)}_{f_4\ldots f_n}\big|^2 = \langle \mathcal{M}_{f_4\ldots f_n}|\{\mathbf{T}_i\cdot \mathbf{T}_k,\mathbf{T}_j
     \cdot \mathbf{T}_l\}|\mathcal{M}_{f_4\ldots f_n}\rangle\,.
\end{equation}
Besides the two- and four-parton correlations, which have already appeared in the context of single- and double-soft emission at tree-level, we also find a non-vanishing three-parton correlation,
\begin{align}
\begin{split}
    \mathscr{S}_{123}^{(\text{trip})}
    |\mathcal{M}_{\bar{q}_1{q}_2g_3f_4\ldots f_n}|^2 &=
    (\mu^{2\eps}g_s^2)^3\ \frac{1}{2}
    \sum_{i,j,k=4}^n\, Q_{ijk}^{\bar{q}qg} 
    |\mathcal{M}^{(ijk)}_{f_4\ldots f_n}|^2\,,
    \label{eq:factorisation_soft_qqg_tripole}
\end{split}
\end{align}
where we defined the triple correlated tree-level squared amplitude by
\begin{align}
\label{eq:SQtrip_CSspace}
\begin{split}
    &|\mathcal{M}_{f_4\ldots f_n}^{(ijk)}|^2 = 
    d^{abc}
    \langle \cM_{f_4\ldots f_n}| \mathbf{T}_i^a \mathbf{T}_j^b \mathbf{T}_k^c | \cM_{f_4\ldots f_n}\rangle \,,
\end{split}
\end{align}
and we have introduced the symmetric structure constant $d^{abc}$,
\begin{align}
    d^{abc} = 2\Tr{\left[\left\{t^a,t^b\right\}t^c\right]}\,.
\end{align}
Colour correlations between three hard partons do not appear in the tree-level soft limits considered in refs.~\cite{BASSETTO1983201,Proceedings:1992fla,Campbell:1997hg,Catani:1999ss,Catani:2019nqv}, and to the best of our knowledge it is the first time that they occur in tree-level currents.

Before we discuss the form of the kinematics functions, we
outline how the colour correlations introduced above come about, and in particular how it occurs that only the totally symmetric structure constant $d^{abc}$ enters the tripole correlations, but there is no correlation between three hard lines connected by the antisymmetric structure constant $f^{abc}$. In order to do so, we analyse the
terms obtained from squaring 
the soft current in eq.~\eqref{eq:qqg_soft_current}.
The square of the $\mathbf{K}$ term in eq.~\eqref{eq:K_current} yields
\begin{align}
\big|\mathbf{K} \big|^2
&=
-{T_F \over 16} \sum_{ijkl} 
\Big[
\big\{\big\{\mathbf{T}_{i}^{a},\, \mathbf{T}_{j}^{b}\big\},\, \big\{\mathbf{T}_{k}^{a},\, \mathbf{T}_{l}^{b}\big\}\big\}
+ (j \leftrightarrow l)
\Big]\,
\mathcal{S}_{j l}(p_3)\,Q^{q\bar{q}}_{i k}(p_1,p_2) \,.
\label{eq:proc1}
\end{align}
Using the identity~\cite{Czakon:2011ve},
\begin{align*}
    &\left\{\left\{\mathbf{T}_i^{a},\mathbf{T}_j^{b}\right\},\left\{\mathbf{T}_k^{a},\mathbf{T}_l^{b}\right\}\right\}
    %+\left\{\left\{\mathbf{T}_i^{a},\mathbf{T}_l^{b}\right\},\left\{\mathbf{T}_k^{a},\mathbf{T}_j^{b}\right\}\right\} 
    +(j\leftrightarrow l) =
    \Big[
    2\{\mathbf{T}_i\cdot\mathbf{T}_k,\mathbf{T}_j\cdot\mathbf{T}_l\} \numberthis\\
    &\qquad
    +\frac{1}{2}C_A\big(
    3\mathbf{T}_i\cdot\mathbf{T}_k(\delta_{il}\delta_{jk}+\delta_{ij}\delta_{kl})
    -4\mathbf{T}_i\cdot\mathbf{T}_j(\delta_{ik}\delta_{il}+\delta_{jk}\delta_{j\ell})
    \big)
    +(i \leftrightarrow k)
    \Big]
    +(j \leftrightarrow l)\,,
\end{align*}
\eqn{eq:proc1} is straightforwardly reduced to two- and four-parton colour correlations.
Next, we examine the crossed term, obtained by contracting $\mathbf{K}$ with $\mathbf{W}$.
The product between $\mathbf{K}$ and the second line of eq.~\eqref{eq:W_current} yields
\begin{align}\label{eq:app-sq-1}
- T_F \sum_{i, j, l}  i f^{abc}\,
\big[\{\mathbf{T}_{i}^{a}, \mathbf{T}_{j}^{b}\},\, \mathbf{T}_l^{c}\big]\, \mathcal{Q}^{(1)}_{i j l}
= - 2T_F\,C_A \sum_{i, j}  \mathbf{T}_{i} \cdot \mathbf{T}_{j}\, (\mathcal{Q}^{(1)}_{i j i} - \mathcal{Q}^{(1)}_{i j j})\,,
\end{align}
where $\mathcal{Q}^{(1)}_{i j l}$ is a function of kinematical invariants,
and where in order to obtain the equality
on the right-hand side,
we used the identity~\cite{Czakon:2011ve},
\begin{align}
    if^{abc}\left[\left\{\mathbf{T}_i^a,\mathbf{T}_j^b\right\},\mathbf{T}_k^c\right] =
    2C_A \left(\mathbf T_i \cdot \mathbf T_j\right)(\delta_{ik}-\delta_{jk})\,.
    \label{eq:dipcol}
\end{align}
We see that \eqn{eq:app-sq-1} is reduced to a two-parton colour correlation.
The product between $\mathbf{K}$ and the first line of eq.~\eqref{eq:W_current} contributes to two- and three-parton correlations.
In particular, the product between $\mathbf{K}$ and the second term in the first line of the $\mathbf{W}$ term has the form,
\begin{align} 
\sum_{i,j,l} &
\Big[
\{\mathbf{T}_{i}^{a}, \mathbf{T}_{j}^{b}\}\,\mathbf{T}_l^{c}\,\operatorname{tr}(t^b t^a t^c)
+ \mathbf{T}_l^{c}\,\{\mathbf{T}_{i}^{a}, \mathbf{T}_{j}^{b}\}\,\operatorname{tr}(t^c t^a t^b)
\Big]\mathcal{Q}^{(2)}_{i j l}
\nonumber\\
&=  
\sum_{i,j,l} 
\bigg(
-{i \over 2}\,T_F\,f^{abc}\, \big[ \{\mathbf{T}_{i}^{a}, \mathbf{T}_{j}^{b}\},\, \mathbf{T}_l^{c}\big]
+ d^{abc}\, \mathbf{T}_{i}^{a} \mathbf{T}_{j}^{b} \mathbf{T}_l^{c}
\bigg)
\mathcal{Q}^{(2)}_{i j l},
\end{align}
where, like $\mathcal{Q}^{(1)}$ in eq.\,\eqref{eq:app-sq-1}, $\mathcal{Q}^{(2)}_{ijl}$ is also a function of kinematical invariants, and we used
\begin{align}\label{}
\operatorname{Tr}\big(t^{a} t^{b} t^{c}\big) = \frac{1}{4}d^{a b c} + \frac{i}{2}\,T_F\, f^{a b c} \,.
\end{align}
The contribution from the first term in the first line of $\mathbf{W}$ has a similar form. 
Note, however, that the product between $\mathbf{K}$ and the first term on the first line of $\mathbf{W}$ has a relative sign difference with respect to the second term on the first line of $\mathbf{W}$. Therefore, the kinematic term of the three-parton correlation \eqref{eq:Q_tripole_symm} inherits a relative minus sign under the exchange of the quark and antiquark labels. Conversely, that sign is cancelled in the $f^{abc}$ term since $f^{abc}$ is antisymmetric. Finally, like in eq.\,\eqref{eq:app-sq-1}, the $f^{abc}$ term is reduced to dipole correlations using again the identity \eqref{eq:dipcol}.
This explains the absence of three-parton colour correlations involving an antisymmetric structure constant.

Let us now discuss the form of the kinematic functions. First we note that, due to the charge conjugation symmetry under exchange of the quark and anti-quark labels, the functions are symmetric under an exchange of $p_1$ and $p_2$. Further, while the four-parton and three-parton correlations in eqs.~(\ref{eq:factorisation_soft_qqg_quad}) and~(\ref{eq:factorisation_soft_qqg_tripole}) and the abelian part of the dipole correlations in eq.~(\ref{eq:factorisation_soft_qqg_dipole}) are insensitive to the mass of the emitters, the non-abelian dipole correlation contains additional terms proportional to the mass of emitters, which we have made explicit in eq.~(\ref{eq:factorisation_soft_qqg_dipole}).
This is in line with the dependence of the triple-soft gluon current on the mass~\cite{Catani:2019nqv}.

The four-parton correlated term in eq.~\eqref{eq:factorisation_soft_qqg_quad} only receives contributions from squaring the function $\mathbf{K}^{a,\mu}$ in eq.~\eqref{eq:K_current}. Therefore, the kinematic function associated to the four-parton correlations separates into the product of two eikonal functions, one associated to the soft gluon and the other to the soft $\bar{q}q$ pair,
\begin{align}
%CHECKED
    Q^{\bar{q}qg}_{ik;j\ell}(p_1,p_2,p_3) = -\frac{1}{2}\Eik{j\ell}{3}Q^{\bar{q}q}_{ik}(p_1,p_2)\,,
    \label{eq:qqg_eikonal_quadrupole}
\end{align}
where the single-soft eikonal function and the soft-$\bar{q}q$ function are given in eqs.~\eqref{eq:eik_g} and~\eqref{eq:eik_qq}.
From the symmetry of the colour operator in eq.~\eqref{eq:quadrupole}, it follows that the kinematic factor in eq.~\eqref{eq:qqg_eikonal_quadrupole} features dipole symmetries under the exchanges $(i\leftrightarrow k)$ or
$(j\leftrightarrow \ell)$, as well as the symmetry under $(i,k) \leftrightarrow (j,\ell)$.

Let us now turn to the kinematic functions appearing in the dipole correlations. They must possess the dipole symmetry $(i\leftrightarrow j)$.
The coefficient of $C_F$ in eq.~\eqref{eq:factorisation_soft_qqg_dipole} is given by
\begin{align}
%CHECKED
\begin{split}
    {Q}^{\bar{q}qg\,(\text{ab})}_{ij}&=
    \frac{8}{\chi_{123}^2\chi_{i(123)}\chi_{j(123)}}\bigg\{
    %%%%%%%%%%%%%%%%%
    \frac{D}{2} \bigg[
    \frac{\chi_{1j}\chi_{3i}}{\chi_{13}}
    -\frac{\chi_{12}\chi_{3i}\chi_{3j}}{\chi_{13}\chi_{23}}
    -\chi_{ij}\left(\frac{\chi_{13}}{\chi_{23}}+1\right)\\
    &+\frac{\chi_{1i}\chi_{3j}+\chi_{1j}\chi_{3i}+\chi_{2i}\chi_{3j}}{\chi_{23}}\bigg]
    %%%%%%%%%%%%%%%%%
    -\frac{\chi_{12}\left(\chi_{12}\chi_{ij}-\chi_{1i}\chi_{2j}-\chi_{1j}\chi_{2i}\right)}{\chi_{13}\chi_{23}}\\
    %%%%%%%%%%%%%%%%%
    &+
    \frac{\chi_{12}\left(\chi_{3j}\chi_{1i}+\chi_{2j}\chi_{3i}+2\chi_{3j}\chi_{3i}\right)}{\chi_{13}\chi_{23}}
    +\chi_{ij}\left(\frac{\chi_{13}-\chi_{12}}{\chi_{23}}-\frac{\chi_{12}}{\chi_{13}}+2\right)
    \\&-\frac{\chi_{1j}\left(\chi_{1i}-\chi_{2i}+2\chi_{3i}\right)}{\chi_{13}}
    -\frac{\chi_{3j}\chi_{i(12)}+\chi_{1j}\left(\chi_{3i}-\chi_{2i}\right)+\chi_{2i}\chi_{j(23)}}{\chi_{23}}
    \bigg\} 
    \\&+ (1\leftrightarrow 2)\,,
\end{split}
\label{eq:qabdip}
\end{align}
with $\chi_{i(123)} = \chi_{1i}+\chi_{2i}+\chi_{3i}$, which is
manifestly symmetric under $(i\leftrightarrow j)$ once the $(1\leftrightarrow 2)$ exchange has been added.
The coefficients of $C_A$ in eq.~\eqref{eq:factorisation_soft_qqg_dipole} are more lengthy, and we provide them in appendix~\ref{app:dipole}.

Finally, the kinematic factor appearing in the three-parton correlations can be written
\iffalse
cast in the form,
%
\begin{align}
%CHECKED
    \begin{split}
        &{Q}^{\bar{q}qg}_{ijk}(p_1,p_2,p_3) \\
        &\,\,\,=
        \frac{4}{\chi_{123}\chi_{i(12)}\chi_{k(123)}\chi_{3j}} \bigg\{
        %%%%%%%%%%%%%%%%%%%%%%%%%%
        %%%% q3 << q12 limit %%%%%
        %%%%%%%%%%%%%%%%%%%%%%%%%%
        \frac{2\chi_{2j}}{\chi_{23}}\left(\chi_{ik}
        -\frac{\chi_{1i}\chi_{2k}+\chi_{1k}\chi_{2i}}{\chi_{12}}\right)
        %%%%%%%%%%%%%%%%%%%%%%%%%%
        %%%%%%%%%% rest %%%%%%%%%%
        %%%%%%%%%%%%%%%%%%%%%%%%%%
        +\chi_{ij}\left[\frac{1}{\chi_{12}}\left(\chi_{1k}-\frac{\chi_{13}\chi_{2k}}{\chi_{23}}\right) \right. \\
        &\,\,\,+\left.\frac{\chi_{3k}}{\chi_{23}}\right] +\frac{\chi_{3i}}{\chi_{23}}\left(\frac{\chi_{1j}\chi_{2k}-\chi_{1k}\chi_{2j}}{\chi_{12}}-\chi_{jk}\right)
        +\frac{1}{\chi_{12}}\left(
        \frac{\chi_{13}\left(\chi_{2i}\chi_{jk}+\chi_{2j}\chi_{ik}\right)}{\chi_{23}}+\chi_{1i}\chi_{jk} \right. \\
        &\,\,\,-\chi_{1j}\chi_{ik}-\left.\frac{\chi_{1i}\chi_{2j}\chi_{3k}+\chi_{1j}\chi_{2i}\chi_{3k}}{\chi_{23}}
        \right)
        \bigg\} 
        -(1\leftrightarrow 2)\,.
    \label{eq:Q_tripole}
    \end{split}
\end{align}
%
Note that, since the three-parton correlation (\ref{eq:SQtrip_CSspace}) is fully symmetric under the exchange of the indices $i$, $j$ and $k$, so is the kinematic factor in eq.~\eqref{eq:Q_tripole}, although 
\eqn{eq:Q_tripole} is not manifestly symmetric in $i$, $j$ and $k$.
We can write eq.~\eqref{eq:Q_tripole}
\fi
in an explicitly symmetric form as follows,
\begin{align*}
    Q_{ijk}^{\bar{q}qg}(p_1,p_2,p_3) &=
    \frac{2}{3\chi_{123}} \bigg\{
    \chi_{ij}\bigg[
    \frac{u_{jk}}{\chi_{k(123)}\chi_{i(12)}}
    +\frac{1}{\chi_{j(123)}}\left(\frac{r_k}{\chi_{i(12)}}+\frac{t_{ik}}{\chi_{k(12)}}\right)
    \bigg]
    + \frac{u_{ij;k}}{\chi_{k(123)}}
    \\ & 
    -\frac{2}{\chi_{12}\chi_{23}\chi_{k(123)}}
    \frac{\chi_{2j}\left(\chi_{1i}\chi_{2k}+\chi_{1k}\chi_{2i}\right)}{\chi_{3j}\chi_{i(12)}}
    +\frac{2\chi_{ij}\chi_{2k}}{\chi_{23}\chi_{3k}}\frac{1}{\chi_{i(12)}\chi_{j(123)}} \\
    & +5\text{ permutations of } (ijk)
    \bigg\}- (1\leftrightarrow 2)\,, \numberthis
    \label{eq:Q_tripole_symm}
\end{align*}
with
\begin{align}
\begin{split}
   r_{k} &=\frac{\chi_{13}\chi_{2k}}{\chi_{12}\chi_{23}\chi_{3k}}-\frac{\chi_{1k}}{\chi_{12}\chi_{3k}}+\frac{1}{\chi_{23}}\,,
   \\t_{ik} &=
   \frac{\chi_{13}\chi_{2k}}{\chi_{12}\chi_{23}\chi_{3i}}+\frac{\chi_{1k}}{\chi_{12}\chi_{3i}}-\frac{\chi_{3k}}{\chi_{23}\chi_{3i}}\,,
   \\u_{ik} &=
   -\frac{\chi_{13}\chi_{2k}}{\chi_{12}\chi_{23}\chi_{3i}}+\frac{\chi_{1k}}{\chi_{12}\chi_{3i}}+\frac{\chi_{3k}}{\chi_{23}\chi_{3i}}\,, \\
   u_{ij;k} &=
   \frac{-\chi_{1i}\left(\chi_{2j}\chi_{3k}+\chi_{2k}\chi_{3j}\right)+\chi_{1j}\left(\chi_{2k}\chi_{3i}-\chi_{2i}\chi_{3k}\right)-\chi_{1k}\left(\chi_{2i}\chi_{3j}+\chi_{2j}\chi_{3i}\right)}{\chi_{12}\chi_{23}\chi_{3j}\chi_{i(12)}}\,.
\end{split}
\end{align}

\subsection{Strongly-ordered soft limits}
\label{sec:strongly_ordered}

In this section we examine the strongly-ordered soft limits, where one or more partons are much softer than the others. 
The strongly-ordered limits can be computed by rescaling the softer momenta by a parameter $\lambda$ and keeping only the leading term in the Laurent expansion around $\lambda=0$. Since the rescaling of the momentum components is uniform in $\lambda$, the ordering in the momentum components is equivalent to an ordering of the energies $E_i$ of the soft particles. We will use the latter to label strongly-ordered limits. We note that the strongly-ordered soft limits where the quark is much softer than the anti-quark (or vice versa) is subleading, and we will not discuss those cases here. Hence, we discuss in the following only the two strongly-ordered soft limits where the energies of the quark and the anti-quark are of the same order.

The factorisation of the squared matrix element in any of the strongly-ordered limits can be written as in eq.~\eqref{eq:qqg_soft}, after replacing the kinematic factors with their expressions in the limit. We now discuss the two relevant strongly-ordered limits in turn.

Let us start with the case where the gluon is softer than the $\bar{q}q$ pair, $E_3 \ll E_1, E_2$. The kinematic functions behave as
    \begin{align*}
       & \mathscr{S}_3Q_{ik;jl}^{\bar{q}qg}(p_1,p_2,p_3)  = Q_{ik;jl}^{\bar{q}qg}(p_1,p_2,p_3)\,,
       \\
            %CHECKED
  &  \mathscr{S}_3\,Q_{ijk}^{\bar{q}qg}(p_1,p_2,p_3) = \frac{1}{6}Q_{ik}^{\bar{q}q}(p_1,p_2)
    \left[\Eik{1j}{3}-\Eik{2j}{3}\right] + 5 \text{ permutations of } (ijk)\,,
    \\
   &             \mathscr{S}_{3}\,{Q}^{\bar{q}qg\,(\text{ab})}_{ij}(p_1,p_2,p_3)  = 2\,\Eik{12}{3}\,Q^{\bar{q}q}_{ij}(p_1,p_2)\,,
   \\
 &   \mathscr{S}_{3}\,{Q}^{\bar{q}qg\,(\text{nab})}_{ij}(p_1,p_2,p_3) =
    %1-line current squared
    \bigg\{
    \frac{1}{4} Q^{\bar{q}q}_{ij}(p_1,p_2)\,\left[
    \Eik{1i}{3}
    %-\frac{1}{4}\Eik{ij}{3} + 2-line =
    -\Eik{ij}{3}
    -\Eik{12}{3}%\frac{2\chi_{12}}{\chi_{13}\chi_{23}}
    \right]\\
    %interference 
&\quad   +\frac{\chi_{12}}{8}\Eik{2j}{3}\,[\Eik{1i}{12}\Eik{2j}{12}-\Eik{2i}{12}\Eik{1j}{12}]\, \left[\Eik{1i}{12}-\Eik{2i}{12}\right] \\
    & \quad- \frac{1}{2\chi_{12}}\,\Eik{ij}{12}\,\Eik{2i}{3}+
    %\frac{1}{4}\Eik{ij}{3}\Eik{1j}{12}\Eik{2j}{12} + 2-line =
    \frac{1}{2}\,\Eik{ij}{3}\,\Eik{1i}{12}\,\Eik{2i}{12} 
    + (i\leftrightarrow j) \bigg\} 
  %\\& \quad 
  + (1\leftrightarrow 2)
    \,.
    \\&
    \mathscr{S}_3\,Q_{ij}^{q\bar{q}g\,(\text{mass})}(p_1,p_2,p_3) =
    m_i^2 \left(\frac{4 \chi_{1j}}{\chi_{12} \chi_{13} \chi_{i(12)}^2 \chi_{3j}}-\frac{4 \chi_{ij}}{\chi_{12}
   \chi_{3i} \chi_{i(12)}^2 \chi_{3j}} + (1\leftrightarrow 2)\right)
   \\&\qquad\qquad\qquad\qquad\qquad + (i\leftrightarrow j)\,. \numberthis
    \end{align*}
    We note that the function $Q_{ik;jl}^{\bar{q}qg}$
    in eq.~\eqref{eq:qqg_eikonal_quadrupole} is exact in this strongly-ordered limit. In the case where the $\bar{q}q$ pair is softer than the gluon, $E_1, E_2 \ll E_3$, the kinematic factors behave as
    \begingroup
    \allowdisplaybreaks
    \begin{align*}
              & \mathscr{S}_{12}Q_{ik;jl}^{\bar{q}qg}(p_1,p_2,p_3)  = Q_{ik;jl}^{\bar{q}qg}(p_1,p_2,p_3)\,,\\ 
         &      \mathscr{S}_{12}\,Q_{ijk}^{\bar{q}qg}(p_1,p_2,p_3)  = \mathscr{S}_{12}\,{Q}^{\bar{q}qg\,(\text{ab})}_{ij}(p_1,p_2,p_3) =0\,,\\
          &\mathscr{S}_{12}\, Q^{\bar{q}qg\, (\text{nab})}_{ij}(p_1,p_2,p_3) =
        \bigg\{
        %1-line squared CHECKED
        \frac{\chi_{13}}{\chi_{12}\chi_{3(12)}}\,\Eik{1i}{3}\,\left[\frac{2\chi_{23}}{\chi_{12}\chi_{3(12)}}-\frac{1}{2}\,\Eik{2i}{12}\right] \\
        &\quad
        %combining with 2-line contribution
        +\Eik{ij}{3}\bigg[\frac{\chi_{13}}{2\chi_{3(12)}} \,\Eik{2i}{12}^2
        +\frac{\chi_{23}}{\chi_{12}\chi_{3(12)}}\,\Eik{1i}{12}
        -\frac{2\chi_{13}\chi_{23}}{\chi_{12}^2\chi_{3(12)}^2}\\
&\quad        %terms combined with 2-line contribution
        +\frac{1}{4}\left(\Eik{1i}{12}\Eik{2i}{12} - Q^{\bar{q}q}_{ij}(p_1,p_2)\right)
        \bigg]
        \\
        % interference %CHECKED
        &\quad -\Eik{ij}{12}\,\bigg[\frac{\chi_{12}}{8\chi_{3(12)}}\,\Eik{1j}{12}\,\Eik{2i}{12}\left(\chi_{13}\,\Eik{1j}{3}-\chi_{23}\,\Eik{i2}{3}\right) \\
        &\quad
        +\frac{1}{2}\Eik{i(12)}{3}\left(\Eik{1j}{12}+
        \frac{1}{\chi_{12}}\right)\bigg]
        + (i\leftrightarrow j)\bigg\}
        + (1\leftrightarrow 2)\,,
        \\&   \mathscr{S}_{12}\,Q_{ij}^{q\bar{q}g\,(\text{mass})}(p_1,p_2,p_3) =
   m_i^2 \bigg(\frac{4 \chi_{3j}}{\chi_{12} \chi_{3(12)} \chi_{3i}^2
   \chi_{j(12)}}
   -\frac{8 \chi_{23} \chi_{1j}}{\chi_{12}^2 \chi_{3(12)} \chi_{3i}^2 \chi_{j(12)}}
   \\&\quad-\frac{4
   \chi_{ij}}{\chi_{12} \chi_{3i}^2 \chi_{i(12)} \chi_{j(12)}}
    +\frac{8 \chi_{1i} \chi_{2j}}{\chi_{12}^2 \chi_{3i}^2 \chi_{i(12)}
   \chi_{j(12)}}+\frac{2}{\chi_{12} \chi_{3(12)} \chi_{3i} \chi_{i(12)}}
   + (1\leftrightarrow 2)\bigg)
   \\&\quad+ (i\leftrightarrow j)\,, \numberthis
    \end{align*}
    \endgroup
    where we defined 
    %$s_{i(jk)}=s_{ij}+s_{ik}$ and
    $\Eik{k(ij)}{l}=\Eik{ik}{l}+\Eik{jk}{l}$.
    The functions $Q_{ik;jl}^{\bar{q}qg}$ in eq.~\eqref{eq:qqg_eikonal_quadrupole} is again exact in this limit.
    The tripole contribution is power-suppressed and can therefore be neglected. The same is true for the coefficient of $C_F$ in the dipole contribution.

\section{Soft limits of the splitting amplitudes}
\label{sec:kinematic}

So far we have only considered soft singularities of on-shell tree-level scattering amplitudes. Soft singularities, however, are not the only kinematic limits in which amplitudes become singular and factorise into universal building blocks, but they also exhibit collinear singularities when two or more massless particles become collinear. These limits need to be taken into account when constructing subtraction schemes for higher order computations. In particular, one is also interested in understanding iterated soft and collinear limits. In this section we discuss the behaviour of scattering amplitudes where a  subset $\bar{q}qg$ of collinear massless particles become soft. Before we do this, we give a short review of collinear factorisation in general.

\input{app_dip_u_func}

\subsection{$m$-parton soft limit of the $m$-collinear splitting amplitudes}
\label{sec:mS_mC}

We start by analysing the special case where all $m$ collinear partons are soft.
This implies that the parent parton must itself be soft, and therefore it must then be a gluon. 
The soft limit of the collinear factorisation in eq.~\eqref{eq:split_def_CSspace} then becomes:
\begin{align}
    \mathscr{S}_{1\ldots m}\,\mathscr{C}_{1\ldots m}\,|\cM_{f_1,\ldots f_n} \rangle =
    \mu^\epsilon g_s\,\mathbf{Sp}_{g f_1 \ldots f_m}\,
    \mathbf{J}_g(\pM{}) \, |\cM_{f_{m+1}\ldots f_n}\rangle\,,
    \label{eq:SC_ampl}
\end{align}
where the soft-gluon current is given in \eqn{eq:single_soft}.
Squaring eq.~\eqref{eq:SC_ampl} and summing over spin and colour indices, we obtain
\begin{align}
\begin{split}
    &\mathscr{S}_{1\ldots m}\mathscr{C}_{1\ldots m}\,|\cM_{f_1\ldots f_n}|^2 \\&\qquad=
    \left(\mu^{2\epsilon}g_s^2\right)^{m+1} \left(\frac{2}{s_{1\ldots m}}\right)^{m-1}
    \sum_{i,j={m+1}}^n \mathcal{S}_{ij\mu\nu}(\pM{})
    \hat{P}_{f_1\ldots f_m}^{\mu\nu}
    |\cM_{f_{m+1}\ldots f_n}^{(ij)}|^2\,,
\end{split}
\label{eq:SC_amplSQ}
\end{align}
where $\hat{P}_{f_1\ldots f_m}^{\mu\nu}$ denotes the polarised splitting amplitude for a gluon to split into $m$ partons $f_1,\ldots, f_m$ defined in eq.~\eqref{eq:P_tensor}, and the sum runs over the hard partons. The soft factor $\mathcal{S}_{ij}^{\mu\nu}$ is given by
\begin{equation}
\mathcal{S}_{ij}^{\mu\nu}(\pM{}) = \frac{p_i^{\mu}\,p_j^{\nu}}{(p_i\cdot \pM{})(p_j\cdot \pM{})}\,.
\end{equation}
Equation~\eqref{eq:SC_amplSQ} generalises straightforwardly the analogous derivation for $m=2$ collinear partons~\cite{Somogyi:2005xz,Czakon:2011ve}.

\subsection{Soft $\bar{q}qg$ limit of tree-level splitting amplitudes}
\label{sec:qqgsoft_of_collinear}

In ref.~\cite{DelDuca:2019ggv}, in the case of a tree-level amplitude of $n$ massless partons, the behaviour of an $m$-parton splitting amplitude was analysed where a single gluon, or a $\bar{q}q$ pair, or two gluons from the collinear set become soft. %while the remaining $(n-m)$ non-collinear partons in the collinear factorisation of the amplitude are all taken to be massless.

We consider now the more general case of a tree amplitude with $(m+r)$ massless and $(n-m-r)$ massive partons.
Firstly, we note that for an $m$-parton splitting amplitude
where a single gluon or a $\bar{q}q$ pair from the collinear set become soft, the kinematic coefficients are the same as the ones of ref.~\cite{DelDuca:2019ggv}, because they are degree-zero homogeneous functions of the $(n-m)$ momenta of the non-collinear partons. Thus, the kinematic coefficients
do not depend on the momenta of the non-collinear partons.
Likewise, for an $m$-parton splitting amplitude where two gluons from the collinear set become soft, the massless pieces of the kinematic coefficients are the same as the ones of ref.~\cite{DelDuca:2019ggv}, while the pieces which are proportional to the squared mass of the  partons~\cite{Czakon:2011ve} are not singular in the collinear limit. Thus, we conclude that in the case of an $m$-parton splitting amplitude where a single gluon, or a $\bar{q}q$ pair, or two gluons from the collinear set become soft, the factorisation formulae are the same as the ones of ref.~\cite{DelDuca:2019ggv} even in the more general case of a tree-level amplitude with massive partons.

We now focus on the behaviour of an $m$-parton splitting amplitude $\hat{P}_{\bar{q} qgf_4\ldots f_m}^{ss'}$
in the limit where a $\bar{q}q$ pair and a gluon from the collinear set become soft. The triple soft limit of the splitting amplitude is defined by performing a rescaling by a small parameter $\lambda_s$ as follows,
\begin{align}
\begin{split}
    &z_i \to \lambda_s z_i\,, \quad \K{i} \to \lambda_s \K{i}\,,\quad
    \chi_{ij} \to \lambda_s^2 \chi_{ij}\,,\quad \chi_{ik} \to \lambda_s \chi_{ik}\,, \\
    &1 \leq i,j \leq 3\,, \quad 3 < k \leq m\,.
\end{split}
\label{eq:tripls}
\end{align}
We expand the ensuing splitting amplitude in $\lambda_s$, and keep only the leading pole, of
$\mathcal O(\lambda_s^{-6})$. We will argue that its coefficient is universal.

In order to obtain the factorisation of the splitting amplitude in the triple soft limit from eq.~\eqref{eq:tripls},
we use the fact that the soft limit and the collinear limit commute. We then start from eq.~\eqref{eq:qqg_soft} and take the collinear limit where partons $1$ through $m$ are collinear
using colour conservation for the hard amplitude,
\begin{align}
    \sum_{j=m+1}^n \mathbf{T}^a_j = -\sum_{j=4}^m \mathbf{T}^a_j\,.
    \label{eq:colour_conservation}
\end{align}
By doing so, we obtain the factorisation formula,
\begin{align}
\label{eq:P_soft_qqg}
 \mathscr{S}_{123}&
\left[\left(\frac{2\mu^{2\eps}g_s^2}{s_{1\ldots m}}\right)^{m-1}
\hat{\mathbf{P}}_{\bar{q} qgf_4\ldots f_m}
\right]
= (\mu^{2\eps}g_s^2)^3
\frac{T_F}{\mathcal{C}_{f}}
\Bigg[\sum_{i,j,k,\ell=4}^m {U}_{ijk\ell;123}^{\bar{q}qg}\,
|\mathbf{Sp}_{ff_4\ldots f_m}^{(ik;j\ell)}|^2\\
&+\sum_{i,j,k=4}^m {U}_{ijk;123}^{\bar{q}qg}\,
|\mathbf{Sp}_{ff_4\ldots f_m}^{(ijk)}|^2
+
\sum_{i,j=4}^m \left(C_F \, {U}_{ij;123}^{\bar{q}qg\,(\text{ab})}
    + C_A\, {U}_{ij;123}^{\bar{q}qg\,(\text{nab})}\right)\,
|\mathbf{Sp}_{ff_4\ldots f_m}^{(ij)}|^2
\Bigg]\,,\nonumber
\end{align}
with the factor $\mathcal{C}_{f}$ defined in \eqn{eq:P_def},
and we introduced the two-, three- and four-parton colour-correlated splitting amplitudes,
\begin{align}
\begin{split}
\label{eq:dipccsplit}
|\mathbf{Sp}_{ff_4\ldots f_m}^{(ij)}|^2 &\equiv
\left[\mathbf{Sp}_{ff_4\ldots f_m}\right]^\dagger
\mathbf{T}_i \cdot \mathbf{T}_j
\,\mathbf{Sp}_{ff_4\ldots f_m}\,, 
\\
|\mathbf{Sp}_{ff_4\ldots f_m}^{(ijk)}|^2 &\equiv
d^{abc} 
\left[\mathbf{Sp}_{ff_4\ldots f_m}\right]^\dagger
\mathbf{T}_i^a\mathbf{T}_j^b\mathbf{T}_k^c
\,\mathbf{Sp}_{ff_4\ldots f_m}\,, 
%\label{eq:tripccsplit}
\\
|\mathbf{Sp}_{ff_4\ldots f_m}^{(ik;j\ell)}|^2&\equiv
\left[\mathbf{Sp}_{ff_4\ldots f_m}\right]^\dagger
\left\{\mathbf{T}_i \cdot \mathbf{T}_k,\mathbf{T}_j \cdot \mathbf{T}_\ell\right\}
\,\mathbf{Sp}_{ff_4\ldots f_m}\,.
%\label{eq:quadccsplit}
\end{split}
\end{align}
The two- and four-parton colour-correlated splitting amplitudes were introduced in ref.~\cite{DelDuca:2019ggv}.
The coefficients of colour-correlated splitting amplitudes are obtained 
by taking the collinear limit of the soft functions in eqs.~\eqref{eq:factorisation_soft_qqg_dipole}, \eqref{eq:factorisation_soft_qqg_quad}
and \eqref{eq:factorisation_soft_qqg_tripole}.

Applying colour conservation in the hard amplitude, eq.~\eqref{eq:colour_conservation}, 
we see that the hard matrix element completely factorises
from the collinear limit of eqs.~\eqref{eq:factorisation_soft_qqg_dipole},
\eqref{eq:factorisation_soft_qqg_quad}
and \eqref{eq:factorisation_soft_qqg_tripole}.
This is due to colour coherence: a cluster of collinear partons acts coherently as one single coloured object, i.e., the hard partons cannot resolve the individual collinear partons, and the colour correlations are in the space of the collinear partons only.
The procedure to compute the contributions of the two- and four-parton colour correlations has been presented in detail in ref.~\cite{DelDuca:2019ggv} in the context of single- and double-soft limits of splitting amplitudes. The computation of the three-parton colour correlations is similar, and so here we content ourselves to simply state the results.
\begin{itemize}
\item
The coefficient multiplying $C_F$ in the two-parton colour-correlated term in eq.~\eqref{eq:P_soft_qqg} is given by
%
%\beq
%    U_{ij;123}^{\bar{q}qg\,(\text{ab})} =
%     Q_{ij}^{\bar{q}qg\,(\text{ab})} +
%     u_{0\,(\text{dip})}^{(\text{ab})}
%    - u_{1\,(\text{dip})}^{(\text{ab})} \,,
%    \label{eq:U_dipab_def} 
%\eeq
%
%
\beq
    U_{ij;123}^{\bar{q}qg\,(\text{ab})} =
     Q_{ij}^{\bar{q}qg\,(\text{ab})} +
     u_{(\text{dip})}^{(\text{ab})} \,,
    \label{eq:U_dipab_def} 
\eeq
where $Q_{ij}^{q\bar{q}g\,(\text{ab})}$ is given in eq.~\eqref{eq:qabdip} and $u_{(\text{dip})}^{(\text{ab})}$
%colour conservation \eqref{eq:colour_conservation} implies that we obtain a minus sign 
collects contributions with at least
one spectator outside of the collinear set,
    \begin{align}
    \begin{split}
    %CHECKED
        u_{(\text{dip})}^{(\text{ab})}
        &= 
        \frac{8}{\chi_{123}^2z_{123}^2}\bigg[
        \frac{z_2(z_1-z_2) -2z_3z_{12}}{\chi_{23}}
        +\frac{z_1z_2\chi_{12}}{\chi_{13}\chi_{23}}
        \left(\frac{z_3(z_3+z_{123})}{z_1z_2}+2\right) \\
        &\qquad\qquad - \frac{ z_1(z_1-z_2 + 2z_3)}{\chi_{13}}
       +\frac{1}{2}D z_3
        \frac{2z_2\chi_{3(12)}-\chi_{12}z_3}{\chi_{13}\chi_{23}}
        \bigg] \\
        & - \bigg\{ \frac{8}{\chi_{123}^2z_{123}s_{i(123)}}\bigg[ z_i\left(\frac{\chi_{13}}{\chi_{23}}
        -\frac{\chi_{12}\chi_{123}}{\chi_{13}\chi_{23}}+2\right)
        +\frac{z_1(s_{2i}-s_{1i}-2s_{3i})}{\chi_{13}} \\
        &\qquad\qquad\qquad +\frac{D}{2} \frac{\chi_{13}(s_{i(12)}z_3-\chi_{3(12)}z_i) + s_{3i}(\chi_{3(12)}z_1-\chi_{12}z_3)}{\chi_{13}\chi_{23}} 
        \\&\qquad\qquad\qquad -\frac{s_{2i}(z_{23}+z_3-z_1) +s_{3i}z_1+s_{1i}z_3}{\chi_{23}} 
        \\ &\qquad\qquad\qquad
        +\frac{\chi_{12}(z_{3}s_{i(13)}+z_{23}s_{3i}+2z_2s_{1i})}{\chi_{13}\chi_{23}}\bigg] + (i\leftrightarrow j)
        \bigg\}  \\&\quad  
        + (1\leftrightarrow 2)\,,
    \end{split}
    \label{eq:U_dipab_def} 
    \end{align}
%
\iffalse
\begin{align}
    %CHECKED
        &u_{1\,(\text{dip})}^{(\text{ab})}
        = \bigg\{
        \frac{8}{\chi_{123}^2z_{123}\chi_{i(123)}}\bigg[ z_i\left(\frac{\chi_{13}}{\chi_{23}}
        -\frac{\chi_{12}\chi_{123}}{\chi_{13}\chi_{23}}+2\right)
        +\frac{z_1(\chi_{2i}-\chi_{1i}-2\chi_{3i})}{\chi_{13}} \nn\\
        &\qquad\qquad +\frac{D}{2} \frac{\chi_{13}(\chi_{i(12)}z_3-\chi_{3(12)}z_i) + \chi_{3i}(\chi_{3(12)}z_1-\chi_{12}z_3)}{\chi_{13}\chi_{23}} 
        \nn \\&\qquad -\frac{\chi_{2i}(z_{23}+z_3-z_1) +\chi_{3i}z_1+\chi_{1i}z_3}{\chi_{23}}
        +\frac{\chi_{12}(z_{3}\chi_{i(13)}+z_{23}\chi_{3i}+2z_2\chi_{1i})}{\chi_{13}\chi_{23}}\bigg]
        \nn \\&\qquad\qquad + (1\leftrightarrow 2) \bigg\}
        + (i\leftrightarrow j) \,,
%        \end{split}
        \label{eq:U_dipabi_def} 
    \end{align}
    \fi
    %
    where we defined $z_{1\ldots k}=z_1+\ldots+z_k$. The function multiplying $C_A$ in the two-parton colour-correlated term in eq.~\eqref{eq:P_soft_qqg} is
    \beq
        U_{ij;123}^{q\bar{q}g\,(\text{nab})} =
        Q_{ij}^{q\bar{q}g\,(\text{nab})} +
        u_{(\text{dip})}^{(\text{nab})} \,,
%        -u_{1\,(\text{dip})}^{(\text{nab})}\,,
    \label{eq:U_dipnab_def}
    \eeq
    where $Q_{ij}^{q\bar{q}g\,(\text{nab})}$ is given in \eqn{eq:Q_qqg_nab}, while $u_{(\text{dip})}^{(\text{nab})}$ is provided in appendix~\ref{app:SC}.
    Note that there is no mass term in \eqn{eq:P_soft_qqg}, since the mass-dependent term $Q_{ij}^{q\bar{q}g\,(\text{mass})}$ in \eqn{eq:Q_qqg_mass} is not singular in the collinear limit.
\item
The function multiplying the three-parton colour correlations in \eqn{eq:P_soft_qqg} reads
\begin{align}
    U_{ijk;123}^{\bar{q}qg} &=
    Q_{ijk}^{\bar{q}qg} +
    u_{(\text{trip})} \,,
%    +u_{1\,(\text{trip})}
%    -u_{2\,(\text{trip})} \,,
\label{eq:U_trip_def}
\end{align}
where $Q_{ijk}^{\bar{q}qg}$ is given in eq.~\eqref{eq:Q_tripole_symm}, while $u_{(\text{trip})}$ is %provided in appendix~\ref{app:SC3}.
\begin{align*}
    u_{(\text{trip})} &=
    \bigg\{
    %%%%%%%%%%%%%%%%%
     \frac{8z_ 1z_ 2z_{23}}{3\chi_{123}\chi_{12}\chi_{23}z_ 3z_{12}z_{123}}
    %%%%%%%%%%%%%%%%%
    +\frac{2}{3\chi_{12}\chi_{123}\chi_{23}}\bigg\{
    \frac{2z_2\left(\chi_{1(23)}z_i-\chi_{1i}z_{23}-z_1\chi_{i(23)}\right)}{z_3z_{12}\chi_{i(123)}}
    \\&+\frac{2}{z_{123}}\bigg[
    \frac{z_1z_2}{z_{12}}
    \left(\frac{\chi_{23}z_i-z_3\chi_{2i}}{z_2\chi_{3i}}-\frac{2\chi_{2i}}{\chi_{3i}}-1\right)
    -\frac{z_{23}\left(-\chi_{12}z_i+z_2\chi_{1i}+z_1\chi_{2i}\right)}{z_3\chi_{i(12)}}
    \bigg]
    %%%%%%%%%%%%%%%%%
    \\&-\chi_{ij}\bigg[
    \frac{z_3\chi_{12}-z_2\chi_{13}+z_1\chi_{23}}{z_{123}\chi_{3i}\chi_{j(12)}}
    +\frac{1}{\chi_{i(123)}}\bigg(\frac{z_2\chi_{13}+z_1\chi_{23}-z_3\chi_{12}}{z_{12}\chi_{3j}}
    \\&+\frac{\left(2z_2+z_3\right)\chi_{12}+z_2\chi_{13}-z_1\chi_{23}}{z_3\chi_{j(12)}}\bigg)
    \bigg]
    -z_i\bigg[
    \frac{1}{\chi_{3i}}\left(\frac{u_{2;j}}{z_{12}\chi_{j(123)}}+\frac{u_{1;j}}{z_{123}\chi_{j(12)}}\right)
    \\&+\frac{1}{\chi_{i(12)}}\left(\frac{\chi_{12}\chi_{3j}+u_{3;j}}{z_{123}\chi_{3j}}+\frac{u_{2;j}}{z_3\chi_{j(123)}}\right)
    +\frac{1}{\chi_{i(123)}}\left(\frac{\chi_{12}\chi_{3j}+u_{3;j}}{z_{12}\chi_{3j}}+\frac{u_{1;j}}{z_3\chi_{j(12)}}\right)
    \bigg]
    \\&-\frac{1}{\chi_{i(123)}}\bigg[
    \frac{\chi_{2i}\left(z_3\chi_{1j}-z_1\left(2\chi_{2j}+\chi_{3j}\right)\right)+r_{ij}}{z_{12}\chi_{3j}}
    +\frac{\chi_{2i}\left(z_1\chi_{3j}-\left(2z_2+z_3\right)\chi_{1j}\right)+r_{ij}}{z_3\chi_{j(12)}}
    \bigg]
    \\&-\frac{z_2\chi_{1j}\chi_{3i}-\chi_{3j}\left(z_2\chi_{1i}+z_1\chi_{2i}\right)-z_1\chi_{2j}\chi_{3i}}{z_{123}\chi_{3j}\chi_{i(12)}}
    \\& +\frac{z_3\chi_{1i}\chi_{2j}+\chi_{2i}\left(\left(2z_2+z_3\right)\chi_{1j}+2z_1\chi_{2j}\right)}{z_{123}\chi_{3i}\chi_{j(12)}}
    \bigg\}
    \\&+ 5 \text{ permutations of } (ijk)
    \bigg\} 
    -(1\leftrightarrow 2)\,,
    \numberthis
\end{align*}
with
\begin{align}
\begin{split}
    u_{1;j}&=-\chi_{12}\chi_{3j}+\chi_{13}\chi_{2j}+\chi_{1j}\chi_{23}\,,\\
    u_{2;j}&=\chi_{12}\chi_{3j}-\chi_{13}\chi_{2j}+\chi_{1j}\chi_{23}\,, \\
    u_{3;j}&=\chi_{2j}\left(2\chi_{12}+\chi_{13}\right)-\chi_{1j}\chi_{23}\,,\\
    r_{ij} &=
    -z_2\left(\chi_{1i}\chi_{3j}+\chi_{1j}\chi_{3i}\right)
    -\chi_{2j}\left(\left(z_2+z_{23}\right)\chi_{1i}+z_1\chi_{3i}\right)\,.
\end{split}
\end{align}
\item
The function multiplying the symmetrised colour correlation in \eqn{eq:P_soft_qqg} is given by
\begin{align}
%CHECKED
    {U}_{ijk\ell;srt}^{\bar{q}qg} = \frac{1}{2} U^{q\bar{q}}_{ik;sr}U_{j\ell;t}\,,
    \label{eq:U_quad}
\end{align}
with
\begin{align}
%CHECKED
    U_{j\ell;t} &= \frac{2z_j}{z_t \chi_{jt}} + \frac{2z_l}{z_t \chi_{lt}} -\Eik{j\ell}{t}\,, \\
    \begin{split}
    U^{\bar{q}q}_{ik;sr} &= \frac{4}{\chi_{sr}^2}\left[\frac{1}{4}\chi_{sr}^2Q^{\bar{q}q}_{ik}(p_s,p_r) 
    -\frac{z_s \chi_{ri}+z_r\chi_{si}-z_i\chi_{sr}}{z_{sr}\chi_{i(sr)}} \right. \\
    &\qquad -\left.\frac{z_s \chi_{rk}+z_r\chi_{sk}-z_k\chi_{sr}}{z_{sr}\chi_{k(sr)}}
    + \frac{2z_sz_r}{z_{sr}^2}
    \right]\,.
    \end{split}
\end{align}
  We note that just like the function (\ref{eq:qqg_eikonal_quadrupole}),
  eq.~\eqref{eq:U_quad} is the product of two functions,
  one associated to the single-soft limit and the other to the soft-$\bar{q}q$ limit of splitting amplitudes introduced in ref.~\cite{DelDuca:2019ggv}.
\end{itemize}
Finally, we note that, just like in the two-soft-gluon case, all the kinematic terms in the factorisation formula \eqref{eq:P_soft_qqg} of the splitting amplitude are independent of whether the embedding tree amplitude contains massive partons or not.

\subsubsection{Soft $\bar{q}qg$ limit of the quadruple collinear splitting amplitudes}
\label{sec:3S_4C}

So far all considerations were completely generic and hold for the soft-$\bar{q}qg$ limit of any splitting amplitude. We now focus on the soft-$\bar{q}qg$ limit 
 within a quadruple collinear splitting $f\to \bar{q} q g f$, and we work out all the color factors explicitly. Equation~\eqref{eq:P_soft_qqg} becomes
\begin{align}
\begin{split}
 \mathscr{S}_{123}
\left[\left(\frac{2\mu^{2\eps}g_s^2}{s_{1234}}\right)^{3}
\hat{\mathbf{P}}_{\bar{q} qgf}
\right]
=&\, (\mu^{2\eps}g_s^2)^3
\frac{1}{\mathcal{C}_{f}}
    \Big[
    \left(C_F\, {U}_{44;123}^{\bar{q}qg\,(\text{ab})}
    + C_A\, {U}_{44;123}^{\bar{q}qg\,(\text{nab})}\right)\,
   \big |\mathbf{Sp}_{ff}^{(44)}\big|^2 
    %\right. 
%    \\ & \left. \qquad\qquad\qquad\qquad
\\
&    +U_{444;123}^{q\bar{q}g}
    \big|\mathbf{Sp}_{ff}^{(444)}\big|^2
    +U_{4444;123}^{q\bar{q}g}
\big|\mathbf{Sp}_{ff}^{(44;44)}\big|^2
    \Big]\,,
\end{split}
\label{eq:quadrupole_SC_cc}
\end{align}
where the kinematic coefficients are given in \eqn{eq:dipccsplit}, and the colour-correlated splitting matrices are defined in \eqn{eq:dipccsplit}, with $\mathbf{Sp}_{ff'} = \delta_{ff'}$ (note that $\mathbf{Sp}_{ff'}$ acts as the identity on both colour and spin indices). We can now evaluate the colour correlations in \eqn{eq:dipccsplit} explicitly. For the dipole and symmetric four-parton correlations, the colour algebra is trivial,
\begin{equation}
    \begin{split}
     \frac{1}{\mathcal{C}_{f}}\,   \big |\mathbf{Sp}_{ff}^{(44)}\big|^2 &\,  =  C_4 \qquad \textrm{~~~and~~~}\qquad
          \frac{1}{\mathcal{C}_{f}}\,   \big|\mathbf{Sp}_{ff}^{(44;44)}\big|^2  =  2\,C_4^2\,,
    \end{split}
\end{equation}
where $C_4$ denotes the Casimir in the representation of the fourth collinear parton, with flavour $f$. For the three-parton correlation we need to distinguish two cases, depending on the flavour $f$. If $f=g$, we have $\left(\mathbf{T}_4^c\right)_{ab}=if^{a_4cb_4}$, and using $d^{abc}=2\Tr{\left(\left\{t^{a},t^{b}\right\}t^{c}\right)}$ and $f^{abc}=-2i\Tr{\left(\left[t^{a},t^{b}\right]t^{c}\right)}$, one can check that the three-parton correlation vanishes. If $f$ is an (anti-)quark instead, $\left(\mathbf{T}_4^c\right)_{ab} = t_{a_4b_4}^c$, we repeatedly use the identity,
    \begin{align}
        t^at^b=\frac{1}{2}\left[\frac{1}{N_c} \delta^{ab}\mathbf{1} +\left(d^{abc}+if^{abc}\right)t^c\right]\,,
    \end{align}
   and  we arrive at
    \begin{align}
    \begin{split}
        d^{abc}\,t^a
    \,t^b
    \,t^c &=
%    \left(4C_F^2 -\frac{3}{2}C_FC_A\right)
    \frac{(N_c^2-1)(N_c^2-4)}{4N_c} \,,
    \end{split}
    \end{align}
    where we used that $d^{aab}=0$ and $d^{abc}f^{abd}=0$, and $d^{abc}d^{abd}=\displaystyle\frac{N_c^2-4}{N_c}\delta^{cd}$. %The colour factors in the second to last line have been calculated using the \textsc{Mathematica} package \textsc{FeynCalc}~\cite{MERTIG1991345,Shtabovenko:2016sxi,Shtabovenko:2020gxv}.
    Hence, we obtain that
    \begin{equation}
        \big|\mathbf{Sp}_{ff}^{(444)}\big|^2 = \delta_{fq}\, \frac{(N_c^2-1)(N_c^2-4)}{4N_c}\,,
    \end{equation}
    where 
    \begin{equation}
        \delta_{fq} = \left\{\begin{array}{ll}
        0\,, & \textrm{ if } f=g\,,\\
        1\,, &\textrm{ otherwise}\,.
        \end{array}\right.
    \end{equation}
    Putting it all together, we see that for $m=4$, eq.~\eqref{eq:quadrupole_SC_cc} can be cast in the compact form,
    \begin{align}\label{eq:P_4_soft_final}
 \mathscr{S}_{123}
\left[\left(\frac{2\mu^{2\eps}g_s^2}{s_{1234}}\right)^{3}
\hat{\mathbf{P}}_{\bar{q} qgf}
\right]
=&\, (\mu^{2\eps}g_s^2)^3
    \bigg[
     C_A^2\, \left({U}_{44;123}^{\bar{q}qg\,(\text{nab})} + 2\,    U_{4444;123}^{q\bar{q}g}\right)\\
\nonumber&\,+C_F\,C_A\, {U}_{44;123}^{\bar{q}qg\,(\text{ab})}    +\delta_{fq}\,\frac{(N_c^2-1)(N_c^2-4)}{4N_c}\,U_{444;123}^{q\bar{q}g}
    \bigg]\,.
\end{align}
Equation~\eqref{eq:P_4_soft_final} describes quadruple splitting amplitudes $\hat{\mathbf{P}}_{\bar{q} qgf}$ in the triple soft limit. We stress that the derivation of eq.~\eqref{eq:P_4_soft_final} only requires the knowledge of the triple soft current of section~\ref{sec:triple_soft}. In particular, we did not need any explicit results for the quadruple splitting amplitudes. We checked, however, that eq.~\eqref{eq:P_4_soft_final} agrees with what we obtain by taking the soft limit of the quadruple collinear splitting amplitudes for $f\in\{g,q,q'\}$ of refs.~\cite{DelDuca:2019ggv,DelDuca:2020vst}.

%%%%%%%%%%%%%%%%%%%%%%%%
%%%%%%%%%%%%%%%%%%%%%%%%

%%%%%%%%%%%%%%%%%%%%%%%%%%%%%%
\section{Conclusions}
\label{sec:conclusion}

In this work we have computed for the first time the tree-level current for the emission of a soft quark-antiquark pair in addition to a gluon. Our result is an operator in the colour space of the hard partons, and it is valid for any number of hard particles, irrespective of their flavour, spin, colour or mass. 

We have also considered the square of the soft current and the colour correlations it induces on the squared matrix elements summed over colour and spin quantum numbers. Remarkably, we find that there are only three different types of colour correlations. These include dipole correlations, which correlate two hard partons and appear already for the emission of a single soft gluon, and symmetrised four-parton correlations that show up in two-gluon emissions. A novel feature of our result is the appearance of tripole correlations involving the totally symmetric structure constant $d^{abc}$. To the best of our knowledge, this is the first time that such colour correlations have been observed in soft limits of tree-level matrix elements. 

If our soft current is used in the context of an N$^3$LO computation, e.g., to build a subtraction scheme at N$^3$LO, it is important to understand also kinematic sublimits, i.e., limits where a subset of unresolved particles develop additional singularities. We have in particular studied the strongly-ordered soft limits in which the gluon is softer than the quark pair (or vice-versa). We have worked out how collinear splitting amplitudes behave if a subset $\bar{q}qg$ of collinear particles becomes soft. This leads to a novel type of universal factorisation of splitting amplitudes, and we have derived the universal building blocks entering this factorisation in detail.  

Our soft current was the last missing ingredient to describe all kinematic infrared singularities that can arise in N$^3$LO computations. This brings the understanding of these infrared limits to the same level as at NNLO, thereby opening the way to the development of subtraction schemes to combine real and virtual corrections at N$^3$LO in the future.

%The tree-level soft $q\bar{q}g$ current will feature in subtraction schemes at ${\rm N^3LO}$ accuracy, and it is the only missing item of the list of infrared currents which are necessary to design subtraction schemes at that accuracy. It is possible to consider several kinematic sub-limits. Some of them may lead to singular regions of the phase-space integration, and thus also be relevant for subtraction schemes. We have considered them in sec.~\ref{sec:kinematic}.

%The completion of the list of infrared currents achieves the first important task, and the most general one, in the design of subtraction schemes at ${\rm N^3LO}$ accuracy. 

\section*{Acknowledgements}
We thank Einan Gardi for useful discussions and for a fruitful collaboration through the early stages of this work.

\appendix

\input{app_dipole}
\input{app_SC}

\input{app_iterated}
\newpage

\bibliographystyle{JHEP}
\bibliography{triple_soft}
\end{document}

%% file: app_dip_u_func.tex
\subsection{Review of collinear factorisation}
\label{app:collfact}
Throughout this section we follow closely the notations and conventions of section~\ref{sec:soft_currents}.
Consider a tree-level scattering amplitude for $n$ partons, and assume that a subset of $m$ masslees particles become collinear to a light-like direction $\pM{}$. The approach to the collinear limit is parametrised by a lightcone decomposition as follows,
\begin{equation} 
\label{eq:collinear_momenta}
p_i^\mu = z_i\pM{\mu} + \K{i}^\mu - \frac{\K{i}^2}{2 z_i}\frac{n^\mu}{\pM{}\cdot n}\,,
\qquad
i = 1,\ldots, m\,,
\end{equation}
where $p_i^2=0$ and $\pM{}\cdot\K{i}=0$. Above, $n^\mu$ is an auxiliary light-like vector such that $n\cdot\K{i} = 0$ and $n\cdot p_{i} \neq 0\neq n\cdot\pM{}$. The longitudinal momentum fractions $z_i$ and the transverse momenta $\K{i}$ can be chosen to satisfy the following constraints,
\begin{align}
\sum_{i=1}^m z_i =1 {\rm~~and~~} \sum_{i=1}^m\K{i}^\mu = 0\,.
\label{eq:constraint}
\end{align}
The collinear limit is performed by introducing a uniform scaling parameter $\lambda_c$ as follows,
\begin{align}
\K{i} \to \lambda_c\, \K{i}\,, \quad 1\le i\le m\,.
\end{align}
Then, we expand the matrix element into a Laurent series around $\lambda_c=0$ and only keep the leading divergent term. In the collinear limit a scattering amplitude factorises\footnote{The factorisation in eq.~\eqref{eq:split_def_CSspace} is valid to all orders in perturbation theory if all the collinear particles are in the final state. For the case where at least one collinear parton is in the initial state, the factorisation in eq.~\eqref{eq:split_def_CSspace} is known to hold only at tree-level~\cite{Catani:2011st}.} as~\cite{Amati:1978by,AMATI197854,Ellis:1978sf},
\begin{align}
    \mathscr{C}_{1\ldots m}\, | \cM_{f_1\ldots f_n}\rangle =
    \mathbf{Sp}_{ff_1\ldots f_m}
    | \cM_{f f_{m+1}\ldots f_n}\rangle\,,
    %| \cM_{f f_{m+1}\ldots f_n}(\tilde{P},p_{m+1},\ldots,p_n)\rangle\,,
    \label{eq:split_def_CSspace}
\end{align}
where $f$ denotes the flavour of the parent particle (note that in QCD $f$ can always be inferred from the flavours $f_1\ldots f_m$ of the collinear partons). The symbol $\mathscr{C}_{1\ldots m}$ indicates that the equality only holds up to terms that are power-suppressed in the collinear limit. 
The scattering amplitude on the right-hand side of eq.~\eqref{eq:split_def_CSspace} is obtained from the original amplitude by replacing the collinear particles with the light-like momentum $\pM{}$.
The quantity ${\bf Sp}$ is called the \textit{splitting matrix}~\cite{Catani:2011st,Catani:2003vu} and only depends on the kinematics and the quantum numbers in the collinear set. 
%The colour and spin dependence of the operator $\mathbf{Sp}_{ff_1\ldots f_m}$ can be made explicit by projecting it onto a basis vector,
%
%\begin{align}
%    {\bf Sp}_{ff_1\ldots f_m}^{c,c_1\ldots c_m;s,s_1\ldots s_m} = 
%    \left(\langle c_1,\ldots,c_m| \otimes \langle s_1,\ldots,s_m|\right) \mathbf{Sp}_{ff_1\ldots f_m} \left(|c\rangle \otimes |s\rangle\right)\,.
%    \label{eq:Sp_def_CSspace}
%\end{align}
%
%Notice that in eq.~\eqref{eq:Sp_def_CSspace} $\mathbf{Sp}$ acts onto the colour index $c$ and spin index $s$ of the parent parton from the left. 

The factorisation in eq.~\eqref{eq:split_def_CSspace} implies that also the squared matrix element must factorise,
\begin{align}
\begin{split}
\mathscr{C}_{1\ldots m}
%\langle 1,\ldots, n | 1,\ldots,n\rangle
&\left|\cM_{f_1\ldots f_n}\right|^2 
= \Bigg(\frac{2\mu^{2\epsilon}\,g_s^2}{s_{1\ldots m}}\Bigg)^{m-1} \,\langle s'|\hat{\mathbf{P}}_{f_1\ldots f_m}|s\rangle\, \langle s |\mathcal{T}_{ff_{m+1}\ldots f_n} |s'\rangle
%(\pM{},p_{m+1},\ldots,p_n)
\,,
\end{split}
\label{eq:factorisation}
\end{align}
where we defined the Mandelstam invariant $s_{1\ldots m} \equiv (p_1+\ldots+p_m)^2$.
The \textit{helicity matrix} $\mathcal{T}_{ff_{m+1}\ldots f_n}$ denotes the square of the reduced amplitude by not summing over the spin of the parent parton,
\begin{align}\label{eq:T_def}
\langle s|\mathcal{T}_{ff_{m+1}\ldots f_n} |s'\rangle \equiv
 \langle \cM_{f f_{m+1}\ldots f_n}| s' \rangle \langle s| \cM_{f f_{m+1}\ldots f_n}\rangle\,.
 %_{\otimes s}
 %\sum_{\substack{(s_{m+1},\ldots,s_n)\\(c,c_{m+1},\ldots,c_n)}}\cM_{ff_{m+1}\ldots f_n}^{c,c_{m+1}\ldots c_n;s,s_{m+1}\ldots s_n}\,\left[\cM_{ff_{m+1}\ldots f_n}^{c,c_{m+1}\ldots c_n;s',s_{m+1}\ldots s_n}\right]^{\ast}\,,
\end{align}
%
%where we have suppressed the momenta on which the amplitude depends. The notation $\langle \ldots \rangle_{ss'}$
%The notation 
%$\langle \cM | \ldots |\cM \rangle_{\otimes s}$ 
%$\langle \ldots \rangle_{\otimes s}$ 
%means that we sum over colour and spin indices implicit to the vector $|\cM\rangle$ as in eq.~\eqref{eq:ampl_squared}, except for the spin indices of the parent parton.
The dependence of the helicity matrix on the lightcone direction $\pM{}$ as well as the momenta of the hard partons is not written explicitly. %The tensorial structure of the factorisation in eq.~\eqref{eq:factorisation} is necessary to correctly capture all spin correlations. 
%The quantitiy $\langle s'|\hat{\mathbf{P}}_{f_1\ldots f_m}|s\rangle$ in eq.~\eqref{eq:factorisation} is the polarised splitting amplitude.
The \textit{(polarised) splitting amplitude} $\hat{\mathbf{P}}$ is related to ${\bf Sp}$ by 
\begin{align}\label{eq:P_def}
\Bigg(\frac{2\mu^{2\epsilon}\,g_s^2}{s_{1\ldots m}}\Bigg)^{m-1} \,
%\hat{P}^{ss'}_{f_1\ldots f_m}
\hat{\mathbf{P}}_{f_1\ldots f_m}=
\frac1{\mathcal{C}_{f}}
%\langle s' |
\left[\mathbf{Sp}_{ff_{1}\ldots f_m}\right]^\dagger 
\mathbf{Sp}_{ff_{1}\ldots f_m}
%| s\rangle
\,,
\end{align}
where $\mathcal{C}_{f}$ is the number of colour degrees of freedom of the parent parton with flavour $f$, i.e., $\mathcal{C}_{g}=N_c^2-1$ for a gluon and $\mathcal{C}_q=N_c$ for a quark. In writing down eq.~\eqref{eq:P_def} we implicitly sum over the spin and colour indices of the collinear partons. Due to colour conservation in the hard amplitude, there are no non-trivial colour correlations, i.e., the operator $\hat{\mathbf{P}}_{f_1\ldots f_m}$ acts as the identity on colour space. However, there can be non-trivial spin correlations.  When the parent parton is a quark, helicity conservation implies that the splitting amplitude is proportional to unity in spin space,
\begin{align}
\langle s' |\hat{\mathbf{P}}_{f_1\ldots,f_m}|s \rangle = \delta^{ss'}\, \hat{{P}}_{f_1\ldots f_m}\,,
\label{eq:unpol}
\end{align}
where the quantity $\hat{{P}}_{f_1\ldots f_m}$ is a scalar. 
In the case where the parent parton is a gluon, it is possible to write the splitting amplitude in terms of Lorentz indices as~\cite{Catani:1999ss},
\begin{align}\label{eq:P_tensor}
\langle s|\hat{\bf{P}}_{f_1\ldots f_m}|s'\rangle = \epsilon_{\mu}^{s}(\pM{},n)^{\ast}\,\epsilon_{\nu}^{s'}(\pM{},n)\,\hat{P}^{\mu\nu}_{f_1\ldots f_m}\,,
\end{align}
with
\begin{align}\hat{P}^{\mu\nu}_{f_1\ldots f_m} = g^{\mu\nu}\,A_{f_1\ldots f_m} + \sum_{i,j=1}^m\,\frac{\K{i}^{\mu}\K{j}^{\nu}}{s_{1\ldots m}}\,B_{ij,f_1\ldots f_m}\,.
\end{align}
The splitting amplitudes for the squared matrix element have been computed at tree-level for the emission of up to four collinear partons~\cite{DelDuca:2019ggv,DelDuca:2020vst}. At one-loop level, they been computed for the emission of up to three collinear partons~\cite{Catani:2003vu,Badger:2015cxa,Czakon:2022fqi}.

So far all considerations were valid in the case where the collinear partons have non-zero energies. In applications it can be interesting to understand how splitting amplitudes behave if a subset of collinear partons are soft.
In ref.~\cite{Somogyi:2005xz} the single- and double-soft limits of triple-collinear tree-level splitting amplitudes were considered. Those soft limits were generalised to a generic set of $m$ collinear partons~\cite{DelDuca:2019ggv}, where $m\ge 2$ in the single-soft limit and $m\ge 3$ in the double-soft limit, and applied to the single- and double-soft limits of quadruple-collinear splitting functions. It was shown in particular that the soft limits of splitting amplitudes involve universal quantities reminiscent of soft currents. In the remainder of this section we extend the analysis of refs.~\cite{Somogyi:2005xz,DelDuca:2019ggv} to certain classes of soft limits of more than two collinear partons.

%% file: app_dipole.tex
\section{Two-parton colour correlated soft function}
\label{app:dipole}

In this appendix we provide the explicit expression for the coefficients of $C_A$ in eq.~\eqref{eq:factorisation_soft_qqg_dipole}, with
\begin{align}
\begin{split}
    {Q}^{\bar{q}qg\,(\text{nab})}_{ij} &= 
   \bigg[ \frac{1}{4}\Eik{ij}{3}\left(
    \Eik{1i}{12}\Eik{2i}{12}
    -\frac{3}{4}Q^{q\bar{q}}_{ij}(p_1,p_2)
    \right) \\
    &+ \frac{1}{\chi_{123}^2\chi_{i(123)}\chi_{j(123)}}\left(
    a_0 + a_1 \chi_{ij}
    +a_2 \chi_{ij}^2
    \right) + (1\leftrightarrow 2)\bigg]
    +
        (i\leftrightarrow j)
            \,,
    \end{split}
    \label{eq:Q_qqg_nab}
\end{align}
and
\begin{align}
\begin{split}
    {Q}^{\bar{q}qg\,(\text{mass})}_{ij} &= m_i^2 \bigg[
    \frac{1}{2\chi_{3i}^2}Q_{ij}^{q\bar{q}}(p_1,p_2)
    -\frac{1}{\chi_{12}\chi_{i(12)}^2}\Eik{ij}{3} \\
    &\qquad + \frac{2}{\chi_{123}\chi_{i(12)}\chi_{j(12)}\chi_{i(123)}\chi_{j(123)}}
    \left(c_0 + c_1 \chi_{ij} \right)
    + (1\leftrightarrow 2)\bigg]
     +   (i\leftrightarrow j)
            \,.
        \end{split}
    \label{eq:Q_qqg_mass}
\end{align}
Like for the coefficient of $C_F$ given in eq.\,\eqref{eq:qabdip}, here we have exploited the dipole symmetry under the exchange of the indices labelling the hard emitters, as well as the charge conjugation symmetry of the $q\bar{q}$ current. On the second lines of eqs.~\eqref{eq:Q_qqg_nab} and \eqref{eq:Q_qqg_mass} we have extracted a pre-factor which captures the overall scaling of $\mathcal{O}(\lambda^{-6})$ in the triple soft limit.

The coefficients $a_0$, $a_1$ and $a_2$ on the right-hand side of eq.~\eqref{eq:Q_qqg_nab} are given by,
\begin{align*}
    a_2&=
        %%%%%%%%%%%%%%%%%%%%
        %%%%%%%%%%%%%%%%%%%%
        \frac{\chi_{3(12)}}{\chi_{i(12)}\chi_{j(12)}}\bigg\{
        \frac{\chi_{12}}{\chi_{3(12)}}\left(\frac{1}{2}-\frac{2\chi_{1i}}{\chi_{3i}}\right)
        +\frac{2\chi_{13}}{\chi_{3(12)}}
        +\frac{1}{\chi_{12}}\left[\chi_{13}
        +\frac{\chi_{3(12)}}{\chi_{3i}}\left(\frac{\chi_{1j}\chi_{i(12)}}{\chi_{3j}}-2\chi_{1i}\right)\right]\\
        &+\frac{1}{\chi_{3i}}\left(\frac{2\chi_{1j}\chi_{i(12)}}{\chi_{3j}}-4\chi_{1i}\right)\bigg\}
        %%%%%%%%%%%%%%%%%%%%
        %%%%%%%%%%%%%%%%%%%%
        +\frac{\chi_{12}}{2\chi_{3i}\chi_{3j}}
        %%%%%%%%%%%%%%%%%%%%
        %%%%%%%%%%%%%%%%%%%%
        \,, \numberthis
\end{align*}
\begingroup
\allowdisplaybreaks
\begin{align*}
    a_1&=
       D +
       \frac{2\chi_{12}^2}{\chi_{13}\chi_{23}}-\frac{2\chi_{12}\chi_{1i}}{\chi_{13}\chi_{3i}}-\frac{\chi_{1j}\chi_{2i}}{\chi_{3i}\chi_{3j}}
       +\frac{\chi_{3(12)}}{\chi_{i(12)}\chi_{j(12)}}\bigg\{
       %%%%%%%%%%%%%%%%%%%%%%%%%%%%%%%%%%%%
        %%%%%%%%%%%%%%%%%%%%%%%%%%%%%%%%%%%%
        \frac{\chi_{12}}{\chi_{3(12)}}\bigg[\chi_{1i}\left(\frac{10\chi_{1j}}{\chi_{13}}+\frac{8\chi_{1j}}{\chi_{23}}\right)
        \\&+\chi_{2i}\left(\frac{9\chi_{1j}}{\chi_{13}}+\frac{9\chi_{1j}}{\chi_{23}}\right)\bigg]
        %%%%%%%%%%%%%%%%%%%%%%%%%%%%%%%%%%%%
        %%%%%%%%%%%%%%%%%%%%%%%%%%%%%%%%%%%%
        -\frac{6\chi_{1i}^2\chi_{1j}}{\chi_{3(12)}\chi_{3i}}
        %%%%%%%%%%%%%%%%%%%%%%%%%%%%%%%%%%%%
        %%%%%%%%%%%%%%%%%%%%%%%%%%%%%%%%%%%%
        -\frac{4\chi_{1j}\chi_{3i}}{\chi_{3(12)}} 
        %%%%%%%%%%%%%%%%%%%%%%%%%%%%%%%%%%%%
        %%%%%%%%%%%%%%%%%%%%%%%%%%%%%%%%%%%%
        -\frac{\chi_{2i}^2}{\chi_{3(12)}\chi_{3i}}\left(\frac{\chi_{13}^2\chi_{1j}}{\chi_{12}\chi_{23}}+\frac{2\chi_{13}\chi_{1j}}{\chi_{23}}+2\chi_{1j}\right)
        %%%%%%%%%%%%%%%%%%%%%%%%%%%%%%%%%%%%
        %%%%%%%%%%%%%%%%%%%%%%%%%%%%%%%%%%%%
        \\&+\frac{\chi_{1j}}{\chi_{12}^2\chi_{3(12)}}\bigg[
        \chi_{2i}\left(\frac{\chi_{1j}(\chi_{13}^2-\chi_{23}^2)}{\chi_{3j}}
        +(\chi_{13}-\chi_{23})^2 -4\chi_{13}\chi_{23}
        \right)
        +4\chi_{23}\chi_{1i}(\chi_{23}-\chi_{13})
        \\&+\frac{\chi_{2i}^2\left(\chi_{23}^2-\chi_{13}^2\right)}{\chi_{3i}}
        \bigg]
        %%%%%%%%%%%%%%%%%%%%%%%%%%%%%%%%%%%%
        %%%%%%%%%%%%%%%%%%%%%%%%%%%%%%%%%%%%
        -\frac{\chi_{1j}}{\chi_{12}\chi_{3(12)}}\bigg[
        \frac{\chi_{2i}}{\chi_{3j}}\left(\frac{4\chi_{13}\chi_{1i}\chi_{2j}}{\chi_{3i}}
        +3\chi_{1j}\chi_{23}\right)
        +\frac{3\chi_{13}\chi_{2i}^2+\chi_{1i}^2\left(4\chi_{13}+10\chi_{23}\right)}{\chi_{3i}}
        \\
        &-\chi_{1i}\left(4\chi_{13}+10\chi_{23}\right)\bigg]
        %%%%%%%%%%%%%%%%%%%%%%%%%%%%%%%%%%%%
        %%%%%%%%%%%%%%%%%%%%%%%%%%%%%%%%%%%%
        +\frac{\chi_{2i}\chi_{1j}}{\chi_{3(12)}}\bigg[
        \frac{6\chi_{13}}{\chi_{23}}
        +\frac{\chi_{23}^2}{\chi_{12}\chi_{13}}
        -\frac{\chi_{1i}}{\chi_{3i}}\left(
        \frac{2\chi_{13}^2}{\chi_{12}\chi_{23}}
        +\frac{4\chi_{13}}{\chi_{23}}
        +10\right) 
        \\&+\frac{6\chi_{23}}{\chi_{13}}
        -\frac{\chi_{1j}}{\chi_{3j}}\left(\frac{\chi_{23}^2}{\chi_{12}\chi_{13}}
        +\frac{2\chi_{23}}{\chi_{13}}
        +2\right)
        +\frac{\chi_{13}^2}{\chi_{12}\chi_{23}}
        +9
        \bigg]
        %%%%%%%%%%%%%%%%%%%%%%%%%%%%%%%%%%%%
        %%%%%%%%%%%%%%%%%%%%%%%%%%%%%%%%%%%%
        -\frac{2\chi_{1i}\chi_{1j}}{\chi_{12}^2\chi_{3i}}\left(
        \frac{\chi_{13}\chi_{2i}\chi_{2j}}{\chi_{3j}} +2\chi_{1i}\chi_{23}
        \right)
        \\&+\frac{1}{\chi_{12}}\left[
        \chi_{2i}\left(
        5\chi_{1j}
        -\frac{\chi_{1i}}{\chi_{3i}}\left(\frac{4\chi_{1j}^2}{\chi_{3j}}+10\chi_{1j}\right)\right) -\frac{2\chi_{1j}^2\chi_{2i}^2}{\chi_{3i}\chi_{3j}}-4\chi_{1j}\chi_{3i
        }
        \right] \\
        %%%%%%%%%%%%%%%%%%%%%%%%%%%%%%%%%%%%
        %%%%%%%%%%%%%%%%%%%%%%%%%%%%%%%%%%%%
        &-\frac{\chi_{3(12)}\chi_{2i}}{\chi_{12}^2\chi_{3i}} \left[\chi_{1i}\left(\frac{2\chi_{1j}^2}{\chi_{3j}}
        +2\chi_{1j}\right)
        +\frac{\chi_{1j}^2\chi_{2i}}{\chi_{3j}}\right]
        %%%%%%%%%%%%%%%%%%%%%%%%%%%%%%%%%%%%
        %%%%%%%%%%%%%%%%%%%%%%%%%%%%%%%%%%%%
        +\frac{\chi_{1i}}{\chi_{3(12)}}\left(
        \frac{2\chi_{1j}\chi_{23}^2}{\chi_{12}\chi_{13}}
        +\frac{8\chi_{1j}\chi_{23}}{\chi_{13}} \right. \\
        &-\left.\frac{2\chi_{1j}\chi_{23}\chi_{i(12)}}{\chi_{3i}\chi_{13}}\left(
        \frac{\chi_{23}}{\chi_{12}}
        +2\right)
        +\frac{4\chi_{13}\chi_{1j}}{\chi_{23}}+10\chi_{1j}\right)\bigg\}
        %%%%%%%%%%%%%%%%%%%%%%%%%%%%%%%%%%%
        %%%%%%%%%%%%%%%%%%%%%%%%%%%%%%%%%%%
        - \frac{\chi_{123}}{\chi_{i(12)}\chi_{j(12)}}\bigg\{
        \frac{\chi_{1j}\chi_{2i}^2}{\chi_{23}\chi_{3i}}
        \\&+\frac{2\chi_{1i}\chi_{1j}\chi_{2i}\chi_{3(12)}}{\chi_{13}\chi_{23}\chi_{3i}}
        +\frac{2\chi_{1i}^2\chi_{1j}}{\chi_{13}\chi_{3i}}
        +\frac{\chi_{1j}^2\chi_{2i}}{\chi_{13}\chi_{3j}}
        -\frac{4\chi_{1i}\chi_{1j}\chi_{2i}\chi_{j(12)}}{\chi_{12}\chi_{3i}\chi_{3j}}
        %%%%%%%%%%%%%%%%%%%%%%%%%%%%%%%%%%%
        \\&+\frac{\chi_{3(12)}}{\chi_{i(12)}\chi_{j(12)}}\bigg[
         \frac{2\chi_{1j}^2\big(2\chi_{1i}\chi_{2i}+\chi_{i(12)}^2\big)}{\chi_{23}\chi_{3(12)}} 
    %%%%%%%%%%%%%%%%%%%%%%%%
    %%%%%%%%%%%%%%%%%%%%%%%%
    +\frac{2\chi_{1j}^2}{\chi_{13}\chi_{3(12)}}\left(\frac{4\chi_{1i}\chi_{2i}\chi_{2j}}{\chi_{1j}}+2\chi_{1i}\chi_{2i}+\chi_{i(12)}^2\right)\\
    %%%%%%%%%%%%%%%%%%%%%%%%
    %%%%%%%%%%%%%%%%%%%%%%%%
    &+\frac{\chi_{1j}\chi_{3i}}{\chi_{13}}\left(
    \frac{\chi_{1j}\left(2\chi_{1i}+\chi_{2i}\right)}{\chi_{3j}}
    \left(\frac{\chi_{2j}}{\chi_{23}}+\frac{\chi_{j(12)}}{\chi_{3(12)}}\right)
    +\frac{2\chi_{1i}\left(\chi_{1j}+2\chi_{2j}\right)+\chi_{1j}\chi_{2i}}{\chi_{23}}
    \right) \\
    %%%%%%%%%%%%%%%%%%%%%%%%
    %%%%%%%%%%%%%%%%%%%%%%%%
    &+\frac{\chi_{1j}\chi_{2i}^2\chi_{3j}}{\chi_{13}\chi_{23}}
    %%%%%%%%%%%%%%%%%%%%%%%%
    %%%%%%%%%%%%%%%%%%%%%%%%
    +\frac{\chi_{2i}^2\chi_{3j}}{\chi_{23}\chi_{3i}}
    \left(\frac{\chi_{1i}\chi_{1j}+\chi_{1j}\chi_{2i}}{\chi_{3(12)}}+\frac{\chi_{1i}\chi_{1j}}{\chi_{13}}\right)\\
    %%%%%%%%%%%%%%%%%%%%%%%%
    %%%%%%%%%%%%%%%%%%%%%%%%
    &-\frac{\chi_{1j}^2\chi_{2i}^2}{\chi_{12}^2}\bigg\{
    2 +\frac{4\chi_{1i}}{\chi_{3i}}\bigg[
    \frac{\chi_{23}}{\chi_{3(12)}}\left(
    \frac{\chi_{1i}}{\chi_{2i}}\left(\frac{\chi_{13}}{\chi_{23}}+4\right)
    +\frac{\chi_{13}}{2\chi_{23}}
    +2+\frac{\chi_{1i}^2}{\chi_{2i}^2}
    +\frac{\chi_{2i}}{2\chi_{1i}}
    \right)\\
    &+\frac{\chi_{2j}}{\chi_{3j}}\left(
    \frac{\chi_{1i}}{\chi_{2i}}
    \left(\frac{\chi_{1j}}{\chi_{2j}}+3\right)
    +\frac{\chi_{1j}}{2\chi_{2j}}+3
    +\frac{\chi_{2i}}{2\chi_{1i}}\right)
    +\frac{2\chi_{1i}\chi_{2j}}{\chi_{1j}\chi_{2i}}
    \bigg] \\
    &+\frac{2\chi_{23}}{\chi_{3(12)}}\left(\frac{\chi_{13}\chi_{1j}+\chi_{2j}\left(4\chi_{13}+\chi_{23}\right)}{\chi_{23}\chi_{3j}}
    +\frac{2\chi_{1i}^2}{\chi_{2i}^2}
    +\frac{4\chi_{1i}}{\chi_{2i}}\right)
    \bigg\}\\
    %%%%%%%%%%%%%%%%%%%%%%%%
    %%%%%%%%%%%%%%%%%%%%%%%%
    &+\frac{2\chi_{1j}^2}{\chi_{12}}\bigg\{
    \frac{\chi_{3i}\chi_{1i}}{\chi_{3(12)}}\left(
    \frac{\chi_{1i}\left(2\chi_{1j}+6\chi_{2j}\right)+\chi_{2i}\left(\chi_{1j}+3\chi_{2j}\right)}{\chi_{1i}\chi_{3j}}+2
    \right) \\
    &+\frac{\chi_{1i}^2\chi_{2i}}{\chi_{3(12)}\chi_{3i}}\bigg[
    \frac{8\chi_{2j}}{\chi_{1j}}+1
    +\frac{\chi_{3j}}{\chi_{1j}}\left(
    \frac{3\chi_{2i}}{\chi_{1i}}+\frac{\chi_{2i}\chi_{i(23)}}{\chi_{1i}^2}
    \right)
    +\frac{\chi_{1i}}{\chi_{2i}}\left(\frac{4\chi_{2j}}{\chi_{1j}}+1\right) \\
    &+\frac{\chi_{2i}}{2\chi_{1i}}
    +\frac{\chi_{2i}^2}{2\chi_{1i}^2}\bigg]
    +\frac{\chi_{1i}\chi_{2i}}{\chi_{3(12)}}\bigg[
    \frac{4\chi_{2j}\chi_{i(23)}}{\chi_{1j}\chi_{2i}}+6
    +\frac{\chi_{1i}}{\chi_{2i}}
    +\frac{\chi_{2i}\chi_{j(12)}}{2\chi_{3j}\chi_{1i}}
    +\frac{\chi_{i(23)}}{\chi_{1i}} \bigg]
    \bigg\}
        \bigg]
        \bigg\}\,,\numberthis
\end{align*}
\endgroup
\begingroup
\allowdisplaybreaks
\begin{align*}
%\begin{split}
%CHECKED
    a_0&=
    \frac{\chi_{3(12)}}{\chi_{i(12)}\chi_{j(12)}}\bigg\{
    \frac{2\chi_{1i}^2\chi_{1j}^2}{\chi_{13}\chi_{3(12)}}\left(\frac{\chi_{23}}{\chi_{12}}+2\right)
    %%%%%%%%%%%%%%%%%%%%%%%%%%%%%%%%%%%%
    %%%%%%%%%%%%%%%%%%%%%%%%%%%%%%%%%%%%
    +\frac{\chi_{1j}\chi_{1i}^2\chi_{2i}^2}{\chi_{12}\chi_{3(12)}\chi_{3i}}\left(\frac{2\chi_{1i}\chi_{23}\chi_{2j}}{\chi_{13}\chi_{2i}^2}+\frac{\chi_{13}\chi_{1j}\chi_{2i}}{\chi_{23}\chi_{1i}^2}\right) \\
    %%%%%%%%%%%%%%%%%%%%%%%%%%%%%%%%%%%%
    %%%%%%%%%%%%%%%%%%%%%%%%%%%%%%%%%%%%
    &+\frac{\chi_{1i}\chi_{1j}^2\chi_{23}\chi_{2i}}{\chi_{12}^2\chi_{3(12)}}\left(\frac{4\chi_{1i}+4\chi_{3i}}{\chi_{2i}}
    -\frac{2\chi_{2i}\chi_{3j}}{\chi_{1i}\chi_{1j}}
    +\frac{2\chi_{3i}}{\chi_{1i}}
    -\frac{8\chi_{2j}}{\chi_{1j}}+8\right) 
    +\frac{\chi_{2i}\chi_{3i}\chi_{1j}^2}{\chi_{13}\chi_{23}}\left(\frac{\chi_{2j}}{\chi_{3j}}-\frac{6\chi_{3j}}{\chi_{1j}}\right) 
    \\&+\frac{2\chi_{1j}\chi_{2i}\left(4\chi_{1i}\chi_{2j}+\chi_{2i}\chi_{3j}\right)}{\chi_{12}^2}
    %%%%%%%%%%%%%%%%%%%%%%%%%%%%%%%%%%%%
    %%%%%%%%%%%%%%%%%%%%%%%%%%%%%%%%%%%%
    +\frac{\chi_{1j}^2}{\chi_{3(12)}}\bigg\{
    \frac{\chi_{2i}\chi_{1i}}{\chi_{12}}\bigg[
    \frac{\chi_{1i}\chi_{23}}{\chi_{13}\chi_{3i}}\left(
    \frac{\chi_{2j}\left(2\chi_{13}^2+4\chi_{23}^2\right)}{\chi_{23}^2\chi_{1j}}+2\right)
    \\&-\frac{4\chi_{13}}{\chi_{23}}\left(
    \frac{\chi_{13}^2+\chi_{23}^2}{\chi_{13}^2} +\frac{\chi_{2j}}{\chi_{1j}}
    \right)
    +\frac{\chi_{3i}}{\chi_{13}\chi_{23}\chi_{1i}}\left(
    \frac{\chi_{2j}\left(\chi_{13}^2+\chi_{23}^2\right)}{\chi_{3j}}
    +3\chi_{23}^2-\chi_{13}^2
    \right)
    \bigg]\\
    &+\frac{2\chi_{1i}\chi_{2i}}{\chi_{13}}\left[
    \frac{\chi_{3i}\left(3\chi_{23}-2\chi_{13}\right)}{2\chi_{23}\chi_{1i}}
    -\frac{5\chi_{13}+3\chi_{23}}{\chi_{23}}-\frac{4\chi_{2j}}{\chi_{1j}}
    \right]
    \bigg\} 
    %%%%%%%%%%%%%%%%%%%%%%%%%%%%%%%%%%%%
    %%%%%%%%%%%%%%%%%%%%%%%%%%%%%%%%%%%%
    +\frac{2\chi_{1i}\chi_{1j}^2\chi_{3i}}{\chi_{13}\chi_{23}}\left(
    \frac{\chi_{2j}}{\chi_{3j}}-\frac{3\chi_{3j}}{\chi_{1j}}
    \right)
    \\&+\frac{2\chi_{1i}\chi_{1j}}{\chi_{3(12)}}\left\{
    \frac{\chi_{1j}\chi_{3i}}{\chi_{12}}\left[
    \frac{\chi_{2j}}{\chi_{3j}}\left(\frac{\chi_{13}}{\chi_{23}}+\frac{\chi_{23}}{\chi_{13}}\right)
    +\frac{3\chi_{2(13)}}{\chi_{13}}-\frac{\chi_{12}}{\chi_{23}}
    \right]
    +\frac{\chi_{2j}\chi_{3i}}{\chi_{12}}\left(
    \frac{2\chi_{13}+\chi_{12}}{\chi_{23}}
    +\frac{\chi_{23}}{\chi_{13}}\right)
    \right\} \\
    %%%%%%%%%%%%%%%%%%%%%%%%%%%%%%%%%%%%
    %%%%%%%%%%%%%%%%%%%%%%%%%%%%%%%%%%%%
    &+\frac{\chi_{1j}^2\chi_{2i}^2}{\chi_{13}\chi_{23}}\left(\frac{\chi_{1i}\chi_{3j}}{\chi_{1j}\chi_{3i}}-6\right)
    +\frac{\chi_{2i}^2}{\chi_{3(12)}}\left[
    \frac{\chi_{1j}\chi_{3j}}{\chi_{12}}\left(\frac{3\chi_{1(23)}}{\chi_{23}}-\frac{\chi_{23}}{\chi_{13}}-\frac{2\chi_{12}}{\chi_{13}}\right) \right. \\
    &+\left.\frac{\chi_{1i}\chi_{1j}\chi_{j(13)}}{\chi_{12}\chi_{3i}}\left(\frac{\chi_{13}}{\chi_{23}}+\frac{\chi_{23}}{\chi_{13}}\right)
    +\frac{\chi_{1j}^2}{\chi_{12}}\left(
    \frac{\chi_{2j}}{\chi_{3j}}\left(\frac{\chi_{13}}{\chi_{23}}+\frac{\chi_{23}}{\chi_{13}}\right)-\frac{3\chi_{13}}{\chi_{23}}-\frac{3\chi_{23}}{\chi_{13}}
    \right)
    +\frac{\chi_{1j}^3\chi_{23}}{\chi_{12}\chi_{13}\chi_{3j}}
    \right]
    %%%%%%%%%%%%%%%%%%%%%%%%%%%%%%%%%%%%
    %%%%%%%%%%%%%%%%%%%%%%%%%%%%%%%%%%%%
    \\&+\frac{\chi_{1j}^2\chi_{2i}^2}{\chi_{12}^2}\left[
    \frac{\chi_{2j}}{\chi_{2i}^2}\left(\frac{4\chi_{1i}^2\chi_{i(12)}}{\chi_{1j}\chi_{3i}}+\frac{4\chi_{1i}\chi_{3i}}{\chi_{3j}}\right)
    +\frac{2\chi_{1i}\chi_{j(13)}}{\chi_{1j}\chi_{3i}}
    +\frac{2\chi_{j(12)}}{\chi_{3j}}
    +\frac{2\chi_{2i}}{\chi_{3i}}
    +\frac{2\chi_{2j}\chi_{3i}}{\chi_{2i}\chi_{3j}}
    -2
    \right]
    %%%%%%%%%%%%%%%%%%%%%%%%%%%%%%%%%%%%
    %%%%%%%%%%%%%%%%%%%%%%%%%%%%%%%%%%%%
    \\&+\frac{\chi_{1j}^2\chi_{2i}^2}{\chi_{12}\chi_{3(12)}}\bigg\{
    \frac{4\chi_{1i}\chi_{j(13)}}{\chi_{1j}\chi_{3i}}
    +\frac{\chi_{1i}\chi_{3i}}{\chi_{2i}^2}\left(
    \frac{8\chi_{2j}}{\chi_{3j}}
    -\frac{6\chi_{j(23)}}{\chi_{1j}}+2+\frac{4\chi_{i(12)}}{\chi_{3i}}
    +\frac{6\chi_{1i}^2\chi_{2j}}{\chi_{1j}\chi_{3i}^2}\right)
    +\frac{3\chi_{2i}}{\chi_{3i}}
    \\&+\frac{3\chi_{j(12)}+\chi_{2j}}{\chi_{3j}}
    +\frac{\chi_{1i}}{\chi_{2i}}\left[\frac{\chi_{1i}}{\chi_{3i}}\left(\frac{10\chi_{2j}}{\chi_{1j}}+2\right)
    +\frac{1}{\chi_{1i}}\left(
    \frac{4\chi_{2j}\chi_{3i}}{\chi_{3j}}
    -\frac{6\chi_{3i}\chi_{3j}}{\chi_{1j}}
    -8\chi_{i(12)}\right)\right]
    \bigg\} \\
    %%%%%%%%%%%%%%%%%%%%%%%%%%%%%%%%%%%%
    %%%%%%%%%%%%%%%%%%%%%%%%%%%%%%%%%%%%
    &-\frac{\chi_{12}\chi_{1j}^2\chi_{2i}^2}{\chi_{13}\chi_{23}\chi_{3(12)}}
    \left[
    \frac{4\chi_{1i}\chi_{3i}}{\chi_{2i}^2}\left(\frac{2\chi_{j(23)}}{\chi_{1j}}+1\right)
    +\frac{2\chi_{3i}}{\chi_{2i}}\left(\frac{4\chi_{3j}}{\chi_{1j}}+1\right)
    +\frac{2\chi_{3j}}{\chi_{1j}}\right]
    \bigg\} 
    %%%%%%%%%%%%%%%%%%%%%%%%
    %%%%%%%%%%%%%%%%%%%%%%%%
    \\&+\frac{D}{2}\bigg[
    \frac{\chi_{23}\left(2\chi_{1i}\chi_{1j}+\chi_{1j}\chi_{2i}\right)}{\chi_{12}\chi_{13}}
    +\frac{\chi_{13}\chi_{1j}\chi_{2i}}{\chi_{12}\chi_{23}}
    +\chi_{3i}\chi_{3j}\left(\frac{2\chi_{12}}{\chi_{13}\chi_{23}}+\frac{1}{\chi_{12}}+\frac{2}{\chi_{13}}\right)\\
    &-\chi_{3i}\left(\frac{2\chi_{1j}\chi_{23}}{\chi_{12}\chi_{13}}+\frac{2\chi_{1j}}{\chi_{12}}+\frac{6\chi_{1j}}{\chi_{13}}+\frac{2\chi_{1j}}{\chi_{23}}\right)
    \bigg]
    %%%%%%%%%%%%%%%%%%%%%%%%
    %%%%%%%%%%%%%%%%%%%%%%%%
    -\frac{4\chi_{12}\chi_{1j}\chi_{2i}}{\chi_{13}\chi_{23}}
    +\frac{\chi_{1j}^2\chi_{2i}}{\chi_{13}\chi_{3j}}
    \\&+\frac{1}{\chi_{3i}}\left(\frac{2\chi_{1i}\chi_{1j}\chi_{2i}}{\chi_{13}}+\frac{\chi_{1j}\chi_{2i}^2}{\chi_{23}}\right)
    %%%%%%%%%%%%%%%%%%%%%%%%%%
    %%%%%%%%%%%%%%%%%%%%%%%%%%
    %%%%%%%%%%%%%%%%%%%%%%%%%%
    %%%%%%%%%%%%%%%%%%%%%%%%%%
    +\frac{\chi_{123}}{\chi_{i(12)}\chi_{j(12)}}\Bigg\{
    \frac{\chi_{1j}^2\chi_{2i}}{\chi_{12}\chi_{3i}}\Bigg[
    \frac{2\chi_{1i}^2}{\chi_{13}}\left(
    1-\frac{\chi_{2j}\left(\chi_{3(12)}+\chi_{23}\right)}{\chi_{23}\chi_{1j}}\right)
    \\&+\frac{\chi_{2i}^2}{\chi_{23}}
    -\frac{2\chi_{1i}^3\chi_{2j}}{\chi_{13}\chi_{1j}\chi_{2i}}
    +\frac{\chi_{1i}\chi_{2i}\chi_{3(12)}}{\chi_{13}\chi_{23}}\Bigg]
    +\frac{\chi_{1j}^3\chi_{2i}^2}{\chi_{12}\chi_{13}\chi_{3j}}\left(\frac{\chi_{2j}\chi_{3(12)}}{\chi_{1j}\chi_{23}}+1\right)
    %%%%%%%%%%%%%%%%%%%%%%%%
    %%%%%%%%%%%%%%%%%%%%%%%%
    \\&+\frac{\chi_{3(12)}}{\chi_{i(12)}\chi_{j(12)}}\bigg\{
    %%%%%%%%%%%%%%%%%%%%%%%%
    %%%%%%%%%%%%%%%%%%%%%%%%
    \frac{2\chi_{1j}^2\chi_{2i}^2\left(\chi_{1j}\chi_{2i}-\chi_{1i}\chi_{2j}\right)}{\chi_{12}\chi_{13}\chi_{23}}
    %%%%%%%%%%%%%%%%%%%%%%%%
    %%%%%%%%%%%%%%%%%%%%%%%%
    +\frac{\chi_{1j}^2\chi_{2i}}{\chi_{12}\chi_{3(12)}}\bigg(
    \frac{\chi_{1i}\chi_{1j}\left(\chi_{2i}-2\chi_{1i}\right)+3\chi_{2j}\left(\chi_{2i}^2-2\chi_{1i}^2\right)}{\chi_{13}} \\
    &+\frac{\chi_{1i}\chi_{1j}\left(2\chi_{1i}+3\chi_{2i}\right)+\chi_{2j}\left(\chi_{2i}^2-2\chi_{1i}^2\right)}{\chi_{23}}\bigg)
    %%%%%%%%%%%%%%%%%%%%%%%%
    %%%%%%%%%%%%%%%%%%%%%%%%
    +\frac{4\chi_{1j}^2\chi_{2i}\chi_{i(12)}\left(\chi_{2i}\chi_{j(12)}-2\chi_{1i}\chi_{2j}\right)}{\chi_{12}^2\chi_{3(12)}}\\
    %%%%%%%%%%%%%%%%%%%%%%%%
    %%%%%%%%%%%%%%%%%%%%%%%%
    &+\frac{4\chi_{1i}^2\chi_{1j}^2\chi_{2i}^2}{\chi_{12}^2\chi_{3(12)}\chi_{3i}}
    \bigg(
    \frac{\chi_{2i}\left(3\chi_{1j}-\chi_{2j}\right)}{2\chi_{1i}}
    +\frac{\chi_{1i}\left(\chi_{1j}-3\chi_{2j}\right)}{\chi_{2i}}
    +\left(3\chi_{1j}-3\chi_{2j}\right)\\
    &+\frac{\chi_{1j}\chi_{2i}^2}{2\chi_{1i}^2}
    -\frac{\chi_{1i}^2\chi_{2j}}{\chi_{2i}^2}\bigg)
    %%%%%%%%%%%%%%%%%%%%%%%%
    %%%%%%%%%%%%%%%%%%%%%%%%
    + \frac{\chi_{1j}^2\chi_{2i}^2\chi_{3j}}{\chi_{12}\chi_{3(12)}\chi_{3i}}
    \left(\chi_{1i}\chi_{i(23)}\left(\frac{1}{\chi_{13}}-\frac{1}{\chi_{23}}\right)+\frac{\chi_{2i}\chi_{3i}}{\chi_{13}}\right)\\
    %%%%%%%%%%%%%%%%%%%%%%%%
    %%%%%%%%%%%%%%%%%%%%%%%%
    &+\frac{2\chi_{1j}^3\chi_{2i}^2\left(\chi_{2i}\left(\chi_{1j}+3\chi_{2j}\right)-\chi_{1i}\chi_{2j}\right)}{\chi_{12}^2\chi_{3(12)}\chi_{3j}}
    +\frac{\chi_{1j}^2\chi_{2j}\chi_{3i}\left(\chi_{13}-\chi_{23}\right)\left(\chi_{1j}\chi_{2i}^2-2\chi_{1i}^2\left(\chi_{1j}+2\chi_{2j}\right)\right)}{\chi_{12}\chi_{13}\chi_{23}\chi_{3(12)}\chi_{3j}} \\
    %%%%%%%%%%%%%%%%%%%%%%%%
    %%%%%%%%%%%%%%%%%%%%%%%%
    &+\frac{\chi_{1j}^2\chi_{3i}\chi_{2i}}{\chi_{12}\chi_{3(12)}}\bigg[
    \frac{\chi_{1i}}{\chi_{23}}\left(\frac{\chi_{2i}\chi_{j(12)}}{\chi_{1i}}-\frac{2\chi_{1i}\chi_{2j}}{\chi_{2i}}+2\chi_{1j}\right)
    -\frac{\chi_{2j}\left(2\chi_{1i}+\chi_{2i}\right)}{\chi_{13}}\bigg]
    %%%%%%%%%%%%%%%%%%%%%%%
    %%%%%%%%%%%%%%%%%%%%%%%
    \bigg\}
    \Bigg\}\,.\numberthis
\end{align*}
\endgroup

The coefficients $c_0$ and $c_1$ on the right-hand side of eq.~\eqref{eq:Q_qqg_mass} are given by,
\begin{align*}
    c_1 &=
    \frac{\chi_{3(12)}}{\chi_{i(12)}\chi_{j(12)}}
    \bigg\{
    \frac{\chi_{2i}\chi_{2j}\chi_{j(12)}}{\chi_{3(12)}\chi_{3i}}\left( 
    \frac{\chi_{1i}^2\chi_{3j}}{\chi_{2i}\chi_{2j}\chi_{3i}}
    +\frac{\chi_{2i}\chi_{3j}}{\chi_{2j}\chi_{3i}}
    +\frac{\chi_{2j}\chi_{3i}}{\chi_{2i}\chi_{3j}}
    +\frac{\chi_{3i}}{\chi_{2i}}
    -\frac{\chi_{3j}}{\chi_{2j}}-2
    \right)
    %%%%%%
    \\&+\frac{\chi_{123}}{\chi_{12}}\frac{\chi_{j(12)}}{\chi_{3j}}\bigg[
    \frac{\chi_{1j}\chi_{2j}}{\chi_{3(12)}}\bigg(
    \frac{2\chi_{1i}\chi_{2i}\chi_{3j}^2}{\chi_{1j}\chi_{2j}\chi_{3i}^2}
    -\frac{2\chi_{1i}\chi_{3j}\chi_{j(12)}}{\chi_{1j}\chi_{2j}\chi_{3i}}
    -\frac{\chi_{1i}\chi_{3j}^2}{\chi_{1j}\chi_{2j}\chi_{3i}}+\frac{\chi_{1j}}{\chi_{2j}}
    \\&-\frac{2\chi_{2i}\chi_{3j}}{\chi_{2j}\chi_{3i}}
    +\frac{\chi_{3j}}{\chi_{2j}}+2
    \bigg)
    +\frac{\chi_{1j}\chi_{2i}\chi_{3j}}{\chi_{123}\chi_{3i}}
    \bigg(
    \frac{\chi_{1i}^2\chi_{j(123)}}{\chi_{1j}\chi_{2i}\chi_{3i}}
    -\frac{\chi_{1i}\chi_{j(12)}^2}{\chi_{1j}\chi_{2i}\chi_{3j}}
    +\frac{2\chi_{1i}\chi_{j(12)}}{\chi_{1j}\chi_{3i}}
    \\&+\frac{\chi_{2j}\chi_{3i}\chi_{j(23)}}{\chi_{1j}\chi_{2i}\chi_{3j}}
    +\frac{\chi_{2i}\chi_{j(23)}}{\chi_{1j}\chi_{3i}}
    -\frac{\chi_{2j}\chi_{j(23)}}{\chi_{1j}\chi_{3j}}
    -\frac{\chi_{1j}}{\chi_{3j}}
    -\frac{\chi_{j(23)}}{\chi_{1j}}
    +\frac{\chi_{2i}}{\chi_{3i}}
    -\frac{2\chi_{2j}}{\chi_{3j}}
    \bigg)
    \bigg]
    \bigg\}
    %%%%%%%%%%%%%%%%%%%
    \\&-\frac{\chi_{j(12)}\left(\chi_{3i}\chi_{j(12)}-\chi_{3j}\chi_{i(12)}\right)}{\chi_{3i}^2\chi_{3j}} \numberthis\,,
\end{align*}
and
\begin{align*}
    c_0 &=
    \frac{\chi_{3(12)}}{\chi_{i(12)}\chi_{j(12)}}
    \bigg\{
    %%%%%%%%%%%%%%
    %%%% rest %%%%
    %%%%%%%%%%%%%%
    \frac{\chi_{1j}\chi_{2j}\chi_{i(12)}\chi_{j(12)}}{\chi_{23}\chi_{3(12)}}\bigg(
    \frac{\chi_{3j}^2\chi_{i(12)}}{\chi_{1j}\chi_{2j}\chi_{3i}}+\frac{\chi_{3j}\chi_{i(123)}\chi_{j(12)}}{\chi_{1j}\chi_{2j}\chi_{3i}}+\frac{3\chi_{2j}}{\chi_{1j}}+4
    \bigg)
    %%%%%%%%%%%%%%%%
    %%%% 1/s12^2 %%%
    %%%%%%%%%%%%%%%%
    \\&-\frac{\chi_{13}\chi_{23}\chi_{i(12)}\chi_{j(12)}}{\chi_{12}^2\chi_{3(12)}} \bigg[
    \frac{\chi_{i(12)}\left(\chi_{123}\chi_{3j}+\chi_{3(12)}\chi_{j(12)}\right)\left(\chi_{1i}\chi_{2j}+\chi_{1j}\chi_{2i}\right)}{\chi_{13}\chi_{23}\chi_{3i}^2}
    \\&+\frac{\chi_{j(123)}}{\chi_{3i}}\left(\frac{\chi_{1j}\left(2\chi_{1i}+\chi_{2i}\right)-\chi_{1i}\chi_{2j}}{\chi_{13}}
    -\frac{\chi_{1j}\chi_{2i}-\chi_{2j}\left(\chi_{1i}+2\chi_{2i}\right)}{\chi_{23}}\right)
    \bigg]
    %%%%%%%%%%%%%%%
    %%%% 1/s12 %%%%
    %%%%%%%%%%%%%%%
   \\&+\frac{\chi_{1j}\chi_{2j}\chi_{i(12)}\chi_{j(12)}}{\chi_{12}\chi_{3(12)}}\bigg[
   \frac{\chi_{13}}{\chi_{23}}\left(\frac{\chi_{2j}}{\chi_{1j}}+1\right)
   +\frac{2\chi_{2j}}{\chi_{1j}}+3
   +\frac{\chi_{1j}\chi_{2(13)}}{\chi_{23}\chi_{2j}}
   \\&+\frac{\chi_{j(12)}^2}{\chi_{1j}\chi_{3j}}\left(
   \frac{\chi_{13}}{\chi_{23}}+\frac{\chi_{1j}}{\chi_{2j}}+2
   \right)
   +\frac{\chi_{3j}}{\chi_{3i}}\bigg(
   \frac{3\chi_{1i}}{\chi_{1j}}+\frac{\chi_{1i}}{\chi_{2j}}
   -\frac{\chi_{13}}{\chi_{23}}\left(\frac{\chi_{1i}}{\chi_{1j}}+\frac{\chi_{2i}}{\chi_{1j}}\right)
   \\& +\frac{2\chi_{2i}}{\chi_{2j}}
     +\frac{2\chi_{2i}}{\chi_{1j}}
   \bigg)
   +\frac{\chi_{1i}\chi_{j(12)}}{\chi_{1j}\chi_{3i}}-\frac{\chi_{1i}\chi_{j(12)}}{\chi_{2j}\chi_{3i}}+\frac{2\chi_{3j}^2\chi_{i(12)}}{\chi_{1j}\chi_{2j}\chi_{3i}}
   -\frac{\chi_{13}\chi_{i(12)}\chi_{j(12)}}{\chi_{1j}\chi_{23}\chi_{3i}}
   \bigg]\bigg\}
   %%%%%%%%%%%%%%%%%%%%%%%%%%
   \\&+\frac{2\chi_{2j}\chi_{j(12)}^2}{\chi_{23}\chi_{3j}}
    %
    %%%%%%%%%%%%%%%%%%%%%%%%
    +\frac{\chi_{j(12)}\left(\chi_{3i}-\chi_{i(12)}\right)\left(\chi_{2(13)}\chi_{3j}-\chi_{2j}\chi_{3(12)}\right)}{\chi_{12}\chi_{23}\chi_{3i}}
    \\&-\frac{\chi_{i(12)}\chi_{j(12)}\left(\chi_{1i}\chi_{2j}+\chi_{1j}\chi_{2i}\right)}{\chi_{12}\chi_{3i}^2}\,. \numberthis
    %\label{eq:c0}
\end{align*}

%% file: app_SC.tex
\section{Collinear limit of the two-parton colour correlation}
\label{app:SC}

In this appendix we consider the $u_{(\text{dip})}^{(\text{nab})}$ term in eq.~\eqref{eq:U_dipnab_def},
    \allowdisplaybreaks
    \begingroup
    \begin{align*}
    %\begin{split}
    %CHECKED
        &u_{(\text{dip})}^{(\text{nab})}
        \\
        &\,\,\,= 
        %%%%%%%%%%%%%%%%%%%%%%%%%%%
        %%%%%%%%%%%%%%%%%%%%%%%%%%%
        \frac{4}{\chi_{123}^2  z_{123}^2}\bigg\{
        \frac{D}{4}\left[
        \frac{z_3-z_{12}}{\chi_{12}}\left(
		\frac{4\chi_{12}z_3-2\chi_{13}z_ 2}{\chi_{23}}
        - z_{12}\left(1+\frac{z_3}{z_3-z_{12}}\right)+2z_ 1
        \right) \right.\\
        &\,\,\,
        +\left.\frac{2z_3(z_1-z_3)}{\chi_{23}}
       -\frac{2z_1z_3}{\chi_{13}}
        +\frac{2\chi_{12}z_3^2}{\chi_{13}\chi_{23}}
        \right] 
        %%%%%%%%%%%%%%%%%%%%%%%%%%%
        %%%%%%%%%%%%%%%%%%%%%%%%%%%
        +\frac{z_1\left(z_1-z_2+2z_3\right)}{\chi_{13}}\\
        %%%%%%%%%%%%%%%%%%%%%%%%%%%
        %%%%%%%%%%%%%%%%%%%%%%%%%%%
        &\,\,\,-\frac{z_3\chi_{12}\left(z_{12}+2z_3\right)}{\chi_{13}\chi_{23}}
        %%%%%%%%%%%%%%%%%%%%%%%%%%%
        %%%%%%%%%%%%%%%%%%%%%%%%%%%
        +\frac{2z_1z_2\chi_{23}}{\chi_{12}^2}\bigg[
        \frac{z_{13}\chi_{13}}{z_1\chi_{23}}\left(\frac{z_{12}}{z_3}+1\right)-\frac{z_3z_1}{z_2z_{12}}
        -\frac{\chi_{3(12)}z_{123}}{2z_{12}\chi_{23}}
        \bigg]\\
        %%%%%%%%%%%%%%%%%%%%%%%%%%%
        %%%%%%%%%%%%%%%%%%%%%%%%%%%
        &\,\,\,+\frac{z_1z_2}{\chi_{23}}\left(\frac{z_1-z_{23}-z_3}{z_{12}}+\frac{z_{23}}{z_1}-\frac{3z_3^2}{z_1z_2} -\frac{4z_2+z_3}{z_2}\right)
        %%%%%%%%%%%%%%%%%%%%%%%%%%%
        %%%%%%%%%%%%%%%%%%%%%%%%%%%
        +\frac{z_1z_2}{\chi_{12}}\bigg[
        \frac{z_{123}+2z_{2}}{z_3}\\
        &\,\,\,+\frac{z_3}{z_{12}}\left(
        2z_1\left(\frac{1}{z_3}-\frac{1}{z_2}\right)-7-\frac{2z_2}{z_1}\right)
        -\frac{z_2-2z_3}{4z_1}+\frac{9\left(z_1+2z_3\right)}{4z_2}-3 \\
        &\,\,\,+\frac{\chi_{13}}{\chi_{23}}\left(\frac{z_1-z_{23}-z_3}{z_{12}}+\frac{z_2}{z_1}+\frac{2z_2}{z_3}-2+\frac{3z_3}{z_1}\right)
        \bigg] 
        %%%%%contributes to q3<<q12 limit %%%%%
        +\frac{2z_1z_2\left(z_2\chi_{13}-z_3\chi_{12}\right)}{z_3\chi_{13}\chi_{23}}
        %%%%%contributes to q12<<q3 limit %%%%%
        \\&\,\,\,+\frac{z_3^2}{\chi_{12}}\left(
        \frac{1}{4}(D-6)
        +\frac{\left(z_2^2\chi_{13}+z_1^2\chi_{23}\right)}{z_3z_{12}\chi_{12}}\right)
        \bigg\}
        %%%%%%%
        \\&\,\,\,- \bigg[
            \frac{4}{\chi_{123}^2z_{123}\chi_{i(123)}}\left(
            c_0 + c_1 z_i + c_2 z_i^2\right)
            +(i\leftrightarrow j)\bigg] 
        + (1\leftrightarrow 2) \,, \numberthis
    %\end{split}
    \end{align*}
    \endgroup
    with
    \begingroup
    \allowdisplaybreaks
    \begin{align*}
        c_0 &=
            (D-6)\frac{z_3\chi_{1i}\chi_{3i}}{2\chi_{12}\chi_{i(12)}}
        %%%%
        +\frac{z_3}{\chi_{i(12)}}\left(
        \frac{\chi_{1i}^2\chi_{23}}{\chi_{12}^2}
        +\frac{z_1^2\chi_{i(12)}\chi_{23}\chi_{3i}}{z_3\chi_{12}^2z_{12}} \right)
        %%%%
        +\frac{z_2\chi_{1i}}{2\chi_{23}}\left(\frac{\chi_{2i}}{\chi_{3i}}+\frac{z_2}{z_3}\right)\nonumber
        \\&+\frac{z_1\chi_{2i}^2}{2\chi_{23}\chi_{3i}}
        %%%%
        +\frac{z_1\chi_{2i}}{\chi_{23}}\left(\frac{z_2}{2z_3}-\frac{2\chi_{12}}{\chi_{13}}\right)
        %%%%
        -\frac{\chi_{123}}{\chi_{12}}\bigg\{
            %%%%%
            \frac{z_1\chi_{2i}}{\chi_{12}}\left(\frac{z_3\chi_{i(12)}}{2z_{12}\chi_{3i}}
            +\frac{z_{12}\chi_{3i}}{2z_3\chi_{i(12)}}
            -1\right) \nonumber
            %%%%%
            \\&-\frac{z_2\left(z_2\chi_{1i}\chi_{i(13)}+z_1\chi_{2i}\chi_{i(23)}\right)}{2z_3\chi_{23}\chi_{i(12)}}
            +\frac{z_1z_2\chi_{1i}\chi_{2i}}{z_{12}\chi_{23}\chi_{i(12)}}\bigg(
            \frac{\chi_{2i}^2}{2\chi_{1i}\chi_{3i}}
            -\frac{z_{13}\chi_{2i}^2}{2z_2\chi_{1i}\chi_{3i}}
            -\frac{z_{13}\chi_{2i}}{2z_2\chi_{3i}}\nonumber
            \\&+\frac{\chi_{2i}}{\chi_{1i}}
            +\frac{\chi_{1i}}{2\chi_{3i}}
            -\frac{z_{23}\chi_{i(12)}}{2z_1\chi_{3i}}
            +\frac{\chi_{2i}}{\chi_{3i}}
            +\frac{z_2}{z_1}+\frac{z_2^2}{2z_1z_3}+\frac{z_1}{2z_3}+\frac{z_2}{z_3}+2
            \bigg)
        \bigg\}\nonumber
        %%%%%%%%%%%%%%%
        \\&+\frac{D}{4}\bigg(
            \frac{\chi_{13}\left(z_2\chi_{i(12)}-z_2\chi_{3i}+\left(z_{12}-z_3\right)\chi_{2i}\right)}{\chi_{12}\chi_{23}}
            +\frac{2z_3\chi_{12}\chi_{3i}}{\chi_{13}\chi_{23}}-\frac{2z_1\chi_{3i}}{\chi_{13}}\nonumber
            \\&-\frac{\left(z_{12}-2z_3\right)\chi_{3i}+z_3\left(\chi_{1i}+3\chi_{2i}\right)}{\chi_{23}}
            -\frac{z_1\chi_{1i}+\left(z_3-z_2\right)\chi_{2i}+z_2\chi_{3i}}{\chi_{12}}\bigg)
        %%%%%%%%%%%%%%%   
          \\&+ \frac{\chi_{1i}\chi_{2i}}{2\chi_{23}}\bigg[
            \frac{z_1z_2\chi_{3i}}{z_3\chi_{1i}\chi_{2i}}
            -\frac{\left(z_{12}+6z_3\right)\chi_{3i}}{\chi_{1i}\chi_{2i}}
            +\frac{z_3}{\chi_{3i}}
            +\frac{2z_{23}+z_3-6z_1}{\chi_{1i}}
            -\frac{4z_2+z_3}{\chi_{2i}}
            %%%%%
            %%%%%
            \\&+\frac{z_1}{z_{12}}\left(
            \frac{z_2}{\chi_{2i}}
            -\frac{z_2}{\chi_{1i}}\left(\frac{\chi_{3i}}{\chi_{2i}}+1\right)
            -\frac{z_3z_2}{z_1\chi_{2i}}\right)
            \bigg]
            %%%%%%%%%%%%%%%%%%%%%%%
            %%%%%%%%%%%%%%%%%%%%%%%
            +\frac{z_1\left(\chi_{1i}-\chi_{2i}+2\chi_{3i}\right)}{\chi_{13}} 
            %%%%%%%%%%%%%%%%%%%%%%%
            %%%%%%%%%%%%%%%%%%%%%%%
            \\&-\frac{\chi_{12}\left(z_3\chi_{i(13)}+z_{23}\chi_{3i}\right)}{\chi_{13}\chi_{23}}
            %%%%%%%%%%%%%%%%%%%%%%%
            %%%%%%%%%%%%%%%%%%%%%%%
            %%%%%%%%%%%%%%%%%%%%%%%
            %%%%%%%%%%%%%%%%%%%%%%%
            +\frac{\chi_{1i}\chi_{2i}}{2\chi_{12}}\Bigg\{
            \frac{\chi_{1i}}{\chi_{3i}}\left(\frac{2z_3-z_1}{\chi_{1i}}+\frac{z_2}{\chi_{2i}}\right)
            %%%%%%%%%%%%%%%%%%%%%%%
            %%%%%%%%%%%%%%%%%%%%%%%
            +\frac{\chi_{13}z_{23}}{\chi_{23}\chi_{3i}}
            %%%%%%%%%%%%%%%%%%%%%%%
            %%%%%%%%%%%%%%%%%%%%%%%
            \\&-\frac{\left(z_3z_{12}+\left(z_1-z_2\right)^2\right)\chi_{123}\chi_{2i}}{z_{12}\chi_{23}\chi_{3i}\chi_{i(12)}}
            %%%%%%%%%%%%%%%%%%%%%%%
            %%%%%%%%%%%%%%%%%%%%%%%
            +\frac{z_1}{\chi_{1i}}\bigg\{
            2+
            \frac{z_2\chi_{13}}{z_3\chi_{23}}
            +\frac{z_2}{z_1}\left(\frac{2\chi_{13}}{\chi_{23}}-1\right)
            +\frac{z_3}{z_1}\left(\frac{3\chi_{13}}{\chi_{23}}+2\right)
            \\&-\frac{3\chi_{13}}{\chi_{23}}
            %%%%
            +\frac{\chi_{3i}}{\chi_{2i}}\bigg[
            \frac{z_2}{z_1}\left(\frac{3\chi_{13}}{\chi_{23}}+2\right)
            +\frac{z_2}{z_3}\left(\frac{\chi_{13}}{\chi_{23}}+2\right)
            -\frac{z_2}{z_{12}}\left(\frac{\chi_{12}+2\chi_{13}}{\chi_{23}}+3\right)
            -\frac{3z_3}{z_1}+3\bigg]
            %%%%
            \\&+\frac{\chi_{2i}}{\chi_{3i}}\left[
            \frac{z_1}{z_{12}}\frac{\chi_{1(23)}}{\chi_{23}}\left(1-\frac{z_2}{z_1}+\frac{2\chi_{23}}{\chi_{1(23)}}\right)
            +\frac{\chi_{3(12)}}{\chi_{23}}\right]
            -\frac{z_2}{z_{12}}\bigg(
            \frac{2\chi_{12}+3\chi_{13}}{\chi_{23}}
            -\frac{z_1\chi_{123}}{z_2\chi_{23}}
            +\frac{2z_{23}}{z_2}
            +\frac{2z_3}{z_1}\bigg)
            \bigg\}
            %%%%%%%%%%%%%%%%%%%%%%%
            %%%%%%%%%%%%%%%%%%%%%%%
            \\&+\frac{z_1z_2}{z_{12}\chi_{3i}}\bigg(
            \frac{z_{23}\chi_{1(23)}}{z_1\chi_{23}}
            +\frac{2z_2}{z_1}
            -\frac{z_{23}\chi_{1(23)}}{z_2\chi_{23}}
            +\frac{z_3(z_2-z_1)}{z_1z_2}
            +\frac{z_1}{z_2}+1\bigg)
            %%%%%%%%%%%%%%%%%%%%%%%
            %%%%%%%%%%%%%%%%%%%%%%%
            +\frac{z_2}{\chi_{2i}}
            \bigg(
            \frac{z_{23}\chi_{12}}{z_{12}\chi_{23}}+\frac{z_1}{z_{12}}
            \\&+\frac{z_2}{z_3}\left(\frac{\chi_{13}}{\chi_{23}}+1\right)-\frac{z_1}{z_3}
            -\frac{2\chi_{13}}{\chi_{23}}
            +\frac{z_3+5z_1}{z_2}
            +3\bigg)
            %%%%%%%%%%%%%%%%%%%%%%%
            %%%%%%%%%%%%%%%%%%%%%%%
            +\frac{z_1}{\chi_{i(12)}}\bigg\{
            \frac{z_2\left(z_1^2-z_2^2\right)\chi_{1i}}{z_1^2z_3\chi_{2i}}
            \\&+\frac{z_{12}^2}{z_1z_3}
            +\frac{z_2\chi_{1i}\left(\chi_{1(23)}-5\chi_{23}\right)}{z_1\chi_{23}\chi_{2i}}
            -\frac{z_2}{z_1}\left(\frac{2\chi_{12}+3\chi_{13}}{\chi_{23}}+2\right)
            -\frac{z_3}{z_1}\left(\frac{\chi_{12}+2\chi_{13}}{\chi_{23}}+3\right)
            \\&+\frac{z_{12}\chi_{3i}}{z_1\chi_{2i}}\left(
            \frac{3z_3}{z_{12}}
            -\frac{z_2}{z_3}\left(\frac{\chi_{1(23)}}{\chi_{23}}+1\right)
            -2\right)
            +\frac{2z_2}{z_{12}}\left(
            \frac{z_2}{z_1}\left(\frac{\chi_{1(23)}}{\chi_{23}}-2\right)
            +\frac{2\chi_{1(23)}}{\chi_{23}}
            -\frac{3z_1}{z_2}-4\right)
            \\&+\frac{z_2\chi_{1i}}{z_1\chi_{3i}}\bigg[
            \frac{z_2z_{12}\chi_{3i}}{z_1z_3\chi_{2i}}
            +\frac{z_1}{z_{12}}\left(\frac{\chi_{1(23)}}{\chi_{23}}+1\right)
            -\frac{z_{23}}{z_{12}}\left(\frac{\chi_{1(23)}}{\chi_{23}}+1\right)
            \bigg]
            +\frac{\chi_{2i}}{\chi_{1i}}
            \bigg[
            \frac{2z_2}{z_{12}}\left(\frac{\chi_{1(23)}}{\chi_{23}}-2\right)
            \\&+\frac{z_{13}\chi_{2i}}{z_{12}\chi_{3i}}\left(
            \frac{z_2}{z_{13}}\left(\frac{\chi_{1(23)}}{\chi_{23}}+1\right)
            -\frac{\chi_{1(23)}}{\chi_{23}}-1\right)
            -\frac{6z_1}{z_{12}}+\frac{\chi_{1(23)}}{\chi_{23}}+\frac{z_{123}}{z_3}
            \\&+\frac{\chi_{3i}}{\chi_{2i}}\left(\frac{\chi_{12}}{\chi_{23}}+\frac{3z_3-2z_{12}}{z_1}\right)
            \bigg] 
            +\frac{\chi_{13}}{\chi_{23}}+2
            \bigg\}
            \Bigg\}
            %%%%%%%%%%%%%%%%%%%%%%%
            %%%%%%%%%%%%%%%%%%%%%%%
            %%%%%%%%%%%%%%%%%%%%%%%
            %%%%%%%%%%%%%%%%%%%%%%%
            +\frac{z_1z_2\chi_{1i}\chi_{2i}}{2\chi_{12}^2}
            \Bigg\{
            %%%%%%%%%%%%
            \frac{2\chi_{13}}{z_1\chi_{1i}}\left(\frac{\chi_{3i}}{\chi_{2i}}-2\right)
            %%%%%%%%%%%%
            \\&+\frac{z_2}{z_{12}}\bigg[
            \frac{4\chi_{13}\chi_{i(12)}}{z_1\chi_{1i}\chi_{2i}}
            -\frac{2}{\chi_{i(12)}}\left(
            \frac{2z_{12}\chi_{3(12)}\chi_{1i}}{z_1z_2\chi_{2i}}
            +\frac{\chi_{13}\chi_{i(12)}\chi_{3i}}{z_1\chi_{1i}\chi_{2i}}
            \right)
            +\frac{\chi_{3(12)}(3\chi_{i(12)}-\chi_{3i})}{z_2\chi_{1i}\chi_{2i}}
            \bigg]
            %%%%%%%%%%%%
            \\&+\frac{2z_{12}\chi_{3(12)}\chi_{1i}}{z_3z_1\chi_{2i}\chi_{i(12)}}
            %%%%%%%%%%%%
            +\frac{z_3z_{12}}{z_1z_2}\bigg\{
            \frac{1}{z_{12}}\left(
            \frac{2\chi_{13}}{\chi_{1i}}+\frac{\chi_{3(12)}}{\chi_{3i}}\right)
            +\frac{1}{\chi_{i(12)}}\bigg[
            \frac{1}{z_3}\left(\frac{4\chi_{13}\chi_{2i}}{\chi_{1i}}+\chi_{3(12)}\right)
            \\&+\frac{z_{12}\chi_{3(12)}}{z_3^2}
            -\frac{1}{z_{12}}\left(
            \frac{2\chi_{13}\chi_{2i}}{\chi_{1i}}+\chi_{3(12)}\right)
            \bigg]
            \bigg\}
            %%%%%%%%%%%%
            +\frac{2\chi_{3(12)}}{z_1z_2}
            \left(\frac{z_1\chi_{2i}}{\chi_{1i}\chi_{3i}}
            -\frac{z_2z_{13}}{z_3\chi_{2i}}
            +\frac{z_1z_2\chi_{3i}}{2z_3\chi_{1i}\chi_{2i}}
            \right)
            %%%%%%%%%%%%
            \\&+\frac{\chi_{123}}{\chi_{i(12)}^2}\bigg\{
            \frac{z_{12}\chi_{1i}\chi_{3i}}{z_1z_3\chi_{2i}}
            -\frac{2\chi_{i(12)}^2}{z_1\chi_{2i}}
            +\frac{z_{123}\chi_{3i}\chi_{i(12)}}{z_3\chi_{1i}\chi_{2i}}\left(
            \frac{\chi_{i(123)}\chi_{1i}}{z_1\chi_{3i}}
            +\frac{\chi_{1i}^2}{z_1\chi_{3i}}
            -\frac{\chi_{1i}}{z_1}
            \right)
            \\&+\frac{z_{123}\chi_{i(12)}}{z_1z_2}\bigg(
            \frac{\chi_{i(12)}}{2\chi_{3i}}
            -\frac{3z_{12}}{2z_3}
            +\frac{1}{2\chi_{i(12)}z_{123}}\left(\frac{z_{12}^2\chi_{3i}}{z_3}+\frac{3z_3\chi_{1i}\chi_{2i}}{\chi_{3i}}\right)
            \bigg)
            +\frac{\chi_{1i}\chi_{2i}}{z_{12}\chi_{3i}}\bigg[
            \frac{z_3\chi_{1i}^2}{z_1\chi_{2i}^2}
            \\&+\frac{z_3\chi_{1i}}{\chi_{2i}}\left(\frac{1}{z_2}
            +\frac{3}{z_1}\right)
            +
            \frac{\chi_{i(12)}^2}{\chi_{1i}\chi_{2i}}\bigg(
            \frac{\chi_{i(123)}}{\chi_{2i}}\left(\frac{z_2-z_3}{z_1}-3\right)
            +\frac{z_{123}\chi_{i(13)}}{z_1\chi_{2i}}
            \bigg)
            \bigg]
            \bigg\}
            \Bigg\}
            %%%%%%%%%%%%%%%%%%%%%%%
            %%%%%%%%%%%%%%%%%%%%%%%
            %%%%%%%%%%%%%%%%%%%%%%%
            %%%%%%%%%%%%%%%%%%%%%%%
            \\&-\frac{\chi_{123}}{4z_{12}\chi_{i(12)}}
            %%%%%%%%%%%%%%
            %%%%%%%%%%%%%%
            \frac{2z_{12}\chi_{2i}\left(\chi_{1i}\left(z_{23}-z_1\right)+z_1\chi_{3i}\right)}{\chi_{123}\chi_{23}}\,,\numberthis
    \end{align*}
    \endgroup
    
    \begingroup
    \allowdisplaybreaks
    \begin{align*}
        c_1 &=
        %%%%%%%%%%%%%%%%%%%%%%%%%
        \frac{D}{2}
        %%%%%%%%%%%%%%
        +\frac{\chi_{123}^2z_{123}\chi_{i(123)}}{4\chi_{12}\chi_{3i}z_3}\left[
        \frac{2}{\chi_{12}}\left(\frac{\chi_{1i}\chi_{2i}}{\chi_{i(12)}^2}
        +\frac{z_1z_2}{z_{12}^2}\right)
        -\frac{3}{\chi_{i(12)}z_{12}} \frac{\chi_{2i}z_1+\chi_{1i}z_2}{\chi_{12}}
        \right]\nonumber
        %%%%%%
        \\&-\frac{z_{12}}{\chi_{i(12)}} \bigg[
        \frac{\chi_{3(12)}\chi_{3i}}{2\chi_{12}z_{12}}
        -\frac{2\chi_{2i}z_1z_2}{z_{12}\chi_{12}}
        \left(
        \frac{(\chi_{13}-\chi_{23})^2}{4z_{12}z_2\chi_{12}}
        -\frac{\chi_{13}\chi_{23}}{z_1z_2\chi_{12}}
        +\frac{\chi_{13}^2}{z_1z_{12}\chi_{12}}
        -\frac{z_3\chi_{3(12)}}{2z_1z_2z_{12}}
        \right)
        \bigg]\nonumber
        %%%%%%
        \\&+
        \frac{\chi_{12}^2}{\chi_{13}\chi_{23}}
        -\frac{\chi_{2i}\chi_{12}}{2\chi_{23}\chi_{3i}}
        -\frac{z_2\chi_{12}}{2z_3\chi_{23}}-\frac{z_2\chi_{1i}}{2z_3\chi_{3i}}
        %%%%%%
        -\frac{\chi_{123}}{\chi_{12}}\bigg\{
        \frac{\chi_{i(12)}}{\chi_{3i}}\left(\frac{z_3}{4z_{12}}-\frac{z_1^2\chi_{23}}{z_{12}^2\chi_{12}}\right)\nonumber
            %%%
            \\&-\frac{\chi_{1i}\chi_{2i}}{\chi_{i(12)}^2}\bigg[
            \frac{2z_1\chi_{1i}\chi_{23}}{z_{12}\chi_{2i}\chi_{12}}
            -\frac{z_{123}}{z_{12}}\left(\frac{\chi_{2i}}{\chi_{1i}}
            +1\right)
            +\frac{z_1^2\chi_{23}}{z_{12}^2\chi_{12}}\left(2+\frac{\chi_{2i}\chi_{3(12)}}{\chi_{1i}\chi_{23}}\right)
            +\frac{z_{12}\chi_{13}\chi_{2i}}{z_3\chi_{1i}\chi_{12}}
            \bigg]\nonumber
            %%%%
            \\&+\frac{z_2\chi_{12}\chi_{i(123)}}{2z_3\chi_{23}\chi_{i(12)}}
            -\frac{\chi_{1i}\chi_{2i}}{2\chi_{3i}\chi_{i(12)}}
            -\frac{\chi_{1i}\chi_{2i}}{z_{12}\chi_{3i}}\bigg[
            \frac{z_1z_2}{z_3\chi_{i(12)}}\left(\left(\frac{\chi_{2i}}{\chi_{1i}}+2\right)+\frac{z_1}{z_2}\right)
            -\frac{\chi_{12}z_{123}}{2\chi_{1i}\chi_{23}}
            \bigg]\nonumber
            \\&-\frac{z_1z_2}{2z_3z_{12}} 
            %%%
            +\frac{\left(z_{123}+z_3\right)\chi_{3i}}{4z_3\chi_{i(12)}}
            \bigg\} 
        %%%%%%%%%%%%%%%%%%%%%%%%%
        +\frac{\chi_{12}}{\chi_{13}}
            -\left(\frac{\chi_{13}}{\chi_{23}}+\frac{1}{2}\right)\left(\frac{\chi_{2i}}{\chi_{3i}}+\frac{z_2}{z_3}\right)
            +\frac{3\chi_{12}+2\chi_{13}}{\chi_{23}}
            %%%
            %%%
            \\&-\frac{\chi_{i(12)}}{4\chi_{3i}}-2
            +\frac{z_1}{2z_{12}}\bigg(
            1-\frac{z_3}{z_1}
            +\frac{z_2\left(\chi_{123}+\chi_{3(12)}\right)}{z_1\chi_{23}}
            - \frac{\chi_{1i}}{\chi_{3i}}\bigg)
            -\frac{\chi_{123}z_{123}}{2z_{12}\chi_{23}}\left(\frac{\chi_{2i}}{\chi_{3i}}+1\right)
        %%%%%%%%%%%%%%%%%%%%%%%
        %%%%%%%%%%%%%%%%%%%%%%%
        %%%%%%%%%%%%%%%%%%%%%%%
        %%%%%%%%%%%%%%%%%%%%%%%
        \\&+\frac{\chi_{1i}\chi_{2i}}{2\chi_{12}}
        \Bigg\{
        \frac{\chi_{13}}{\chi_{1i}\chi_{2i}}
            \bigg\{\frac{1}{z_{12}}\left[
            z_1\left(\frac{2\chi_{23}}{\chi_{13}}+1\right)+z_2\left(\frac{\chi_{13}}{\chi_{23}}+2\right)
            -\frac{\chi_{12}z_{23}+2z_3\chi_{3(12)}}{\chi_{13}}
            \right]
            \\&-\frac{z_2}{z_3}\left(\frac{\chi_{13}}{\chi_{23}}
            +\frac{z_1\chi_{23}}{z_2\chi_{13}}+2\right)
            -4\bigg\}
            -\frac{\chi_{13}}{\chi_{1i}\chi_{3i}}
            \bigg(\frac{z_2\chi_{3(12)}+z_{23}\chi_{123}}{z_{12}\chi_{13}}
            +\frac{\chi_{13}}{\chi_{23}}
            \\&+\frac{z_1\chi_{3(12)}}{z_3\chi_{13}}
            +2\bigg)
            +\frac{\chi_{3(12)}}{\chi_{1i}\chi_{i(12)}}
            \bigg\{
            \frac{\chi_{12}}{\chi_{3(12)}}\left[
            \frac{\chi_{13}}{\chi_{12}}\left(\frac{\chi_{13}}{\chi_{23}}+2\right)
            -\frac{z_{12}}{z_3}-1\right]
            \\&+\frac{z_{123}+z_2}{z_{12}}
            -\frac{z_{12}+z_2}{z_3}\bigg\}
            +\frac{\chi_{3(12)}}{\chi_{2i}\chi_{i(12)}}
            \left(\frac{z_{123}+z_1}{z_{12}}+\frac{\chi_{23}}{\chi_{3(12)}}-\frac{z_1}{z_3}+1\right)
            \\&-\frac{\chi_{3(12)}}{\chi_{2i}\chi_{3i}}\bigg(
            \frac{z_{123}}{z_{12}}
            +\frac{z_2}{z_3}
            +\frac{z_{23}\chi_{12}}{z_{12}\chi_{3(12)}}
            +\frac{\chi_{23}}{\chi_{3(12)}}\bigg)
            %%%%
            %%%%
            +\frac{\chi_{123}}{z_{12}z_3}\bigg\{
            \frac{z_1z_2\left(z_{12}-z_3\right)\chi_{i(123)}}{z_{12}\chi_{1i}\chi_{2i}\chi_{3i}}
            \\&-\frac{z_1z_2}{\chi_{3i}\chi_{i(12)}}\bigg[
            \frac{z_{12}^2}{z_1z_2}\left(\frac{\chi_{3i}^2}{\chi_{1i}\chi_{2i}}-1\right)
            +\frac{\chi_{3i}\chi_{i(12)}}{\chi_{1i}\chi_{2i}}
            +\frac{\chi_{2i}}{\chi_{1i}}
            +\frac{\chi_{1i}}{\chi_{2i}}
            -\frac{3z_3^2}{2z_1z_2}
            +\frac{z_2}{z_1}
            +\frac{z_1}{z_2}
            \\&+\frac{z_3z_{12}\chi_{3i}^2}{z_1z_2\chi_{1i}\chi_{2i}}
            +4
            \bigg]
            +\frac{z_{123}\chi_{3i}}{\chi_{i(12)}^2}
            \bigg[
            \frac{z_3^2\chi_{1i}}{z_{123}\chi_{3i}^2}
            \left(\frac{\chi_{1i}}{2\chi_{2i}}+\frac{\chi_{2i}^2}{2\chi_{1i}^2}\right)
            +\frac{z_{12}^2}{z_{123}\chi_{3i}}\left(
            \frac{\chi_{i(12)}\chi_{3i}}{2\chi_{1i}\chi_{2i}}-1\right)
            \\&+\frac{z_{12}z_3}{z_{123}\chi_{3i}}
            \left(\frac{\chi_{i(12)}\chi_{3i}}{\chi_{1i}\chi_{2i}}-1\right)
            +\frac{z_3}{\chi_{3i}}\left(\frac{\chi_{1i}}{\chi_{2i}}+\frac{\chi_{2i}}{\chi_{1i}}+2\right)
            \bigg]
            \bigg\}
        \Bigg\}
        %%%%%%%%%%%%%%%%%%%%%%%
        %%%%%%%%%%%%%%%%%%%%%%%
        %%%%%%%%%%%%%%%%%%%%%%%
        %%%%%%%%%%%%%%%%%%%%%%%
        \\&+\frac{z_1z_2\chi_{1i}\chi_{2i}}{2\chi_{12}^2}
        \bigg\{
        \frac{\chi_{3(12)}}{z_1z_2}\left[
            \left(\frac{\chi_{13}}{\chi_{1i}}+\frac{\chi_{23}}{\chi_{2i}}\right)
            \left(\frac{1}{\chi_{i(12)}}-\frac{1}{\chi_{3i}}\right)
            -\frac{4\chi_{13}\chi_{23}}{\chi_{3(12)}\chi_{1i}\chi_{2i}}
            \right]
            %%%%%%%%%%%%
            \\&+\frac{2\chi_{13}\chi_{3(12)}}{z_1\chi_{1i}\chi_{2i}}\left(\frac{1}{z_{12}}-\frac{1}{z_3}\right)
            %%%%%%%%%%%%
            +\frac{\chi_{3(12)}^2}{z_1\chi_{2i}}\left(\frac{1}{z_{12}\chi_{3i}}-\frac{1}{z_3\chi_{3i}}+\frac{1}{z_3\chi_{i(12)}}\right)
            %%%%%%%%%%%%
            \\&+\frac{2\left(\chi_{13}-\chi_{23}\right)}{z_{12}\chi_{2i}\chi_{i(12)}}\left(\frac{\chi_{23}-\chi_{13}}{z_1}+\frac{2\chi_{23}}{z_2}\right)
            %%%%%%%%%%%%
            %%%%%%%%%%%%
            +\frac{\chi_{123}}{\chi_{i(12)}^2}\bigg\{
            %%%%%%%%%%%%%
            \frac{2\chi_{13}}{z_{12}z_3}\bigg[
            \frac{\chi_{i(12)}^2\chi_{i(123)}}{\chi_{1i}\chi_{2i}\chi_{3i}}\left(
            \frac{z_3}{z_1}
            +\frac{\chi_{3(12)}}{2\chi_{13}}\right)
            \\&-\frac{2z_3\chi_{2i}}{z_1\chi_{1i}}-\frac{3\chi_{2i}}{\chi_{1i}}
            \bigg]
            %%%%%%%%%%%%%
            -\frac{\chi_{3(12)}}{z_{12}^2}\bigg[
            \frac{2z_1\chi_{2i}}{z_2\chi_{1i}}
            +\frac{\chi_{i(12)}^2\chi_{i(123)}}{\chi_{1i}\chi_{2i}\chi_{3i}}
            +\frac{6z_1\chi_{23}\chi_{i(12)}}{z_2\chi_{3(12)}\chi_{3i}}
            \\&+\frac{2\chi_{23}}{\chi_{3(12)}}
            \bigg(
            \frac{z_1\chi_{1i}}{z_2\chi_{3i}}\left(\frac{\chi_{1i}}{\chi_{2i}}
            +\frac{\chi_{2i}^2}{\chi_{1i}^2}\right)
            +\frac{z_1^2\chi_{1i}}{z_2z_3\chi_{2i}}
            +\frac{z_2^2\chi_{1i}}{z_3z_1\chi_{2i}}
            +\frac{2z_1}{z_2}\bigg)
            \bigg]
            %%%%%%%%%%%%
            \\&+\frac{z_{123}\chi_{3(12)}}{z_1z_2z_3}
            \frac{\chi_{i(12)}}{\chi_{3i}}
            \left[
            \frac{\chi_{3i}}{\chi_{3(12)}}\left(
            \frac{\chi_{13}}{\chi_{1i}}+\frac{\chi_{23}}{\chi_{2i}}\right)-\frac{\chi_{3i}}{\chi_{i(12)}}+1\right]
            %%%%%%%%%%%%
            %%%%%%%%%%%%
            \bigg\}
        \Bigg\}
        %%%%%%%%%%%%%%%%%%%%%%%
        %%%%%%%%%%%%%%%%%%%%%%%
        %%%%%%%%%%%%%%%%%%%%%%%
        %%%%%%%%%%%%%%%%%%%%%%%
        \\&+\frac{\chi_{123}}{4z_{12}\chi_{i(12)}}
        \bigg[
        \frac{\chi_{1i}\chi_{2i}}{\chi_{123}}\bigg(
        \frac{30z_1+28z_2}{\chi_{2i}}
        +\frac{2z_{12}\left(\chi_{12}+2\chi_{13}\right)}{\chi_{1i}\chi_{23}}
        -\frac{2z_{12}\chi_{3i}}{\chi_{1i}\chi_{2i}}
        -\frac{2\left(2z_1+z_2\right)z_{12}}{z_3\chi_{2i}}
        \bigg)
        \\&+\frac{2z_{123}\chi_{2i}\chi_{i(12)}}{\chi_{23}\chi_{3i}}
        -\frac{2z_{12}\chi_{i(123)}}{\chi_{23}}
        \bigg]\,,\numberthis
    \end{align*}
    \endgroup
    
    \begingroup
    \allowdisplaybreaks
    \begin{align*}
        c_2 &=
        \frac{\chi_{12}}{4z_3\chi_{3i}}
        +\frac{\chi_{3(12)}^2}{4\chi_{12}z_{12}\chi_{i(12)}}\left(\frac{3\chi_{123}^2z_{123}\chi_{i(123)}}{z_3\chi_{3i}\chi_{3(12)}^2}+1\right)
        %%%%%%%%%%%%%%%%%%%%%%%
        %%%%%%%%%%%%%%%%%%%%%%%
        %%%%%%%%%%%%%%%%%%%%%%%
        %%%%%%%%%%%%%%%%%%%%%%%
        +\frac{\chi_{12}}{2z_{12}\chi_{3i}}\left[
        \frac{\chi_{3(12)}}{\chi_{12}}\left(\frac{z_{12}}{z_3}-1\right)-\frac{1}{2}\right]
        %%%%%%%%%%%%%%%%%%%%%%%
        %%%%%%%%%%%%%%%%%%%%%%%
        %%%%%%%%%%%%%%%%%%%%%%%
        %%%%%%%%%%%%%%%%%%%%%%%
        \\&+\frac{\chi_{1i}\chi_{2i}}{2\chi_{12}}
        \bigg\{
        \frac{\chi_{3(12)}^2}{2z_{12}\chi_{1i}\chi_{2i}\chi_{i(12)}}\left[
        \frac{z_{12}}{z_3}\left(\frac{\chi_{i(12)}}{\chi_{3i}}-1\right)
        -\frac{\chi_{i(12)}}{\chi_{3i}}\right]
        \bigg\}
        %%%%%%%%%%%%%%%%%%%%%%%
        %%%%%%%%%%%%%%%%%%%%%%%
        %%%%%%%%%%%%%%%%%%%%%%%
        %%%%%%%%%%%%%%%%%%%%%%%
        \\&+\frac{\chi_{123}}{4z_{12}\chi_{i(12)}}
        \bigg(
        \frac{\chi_{3(12)}}{\chi_{123}}
        -\frac{z_{12}\left(\chi_{12}+2\chi_{3(12)}\right)}{z_3\chi_{123}}+1
        \bigg)\,. 
        \numberthis
    \end{align*}
    \endgroup